\documentclass[]{iopart}
\usepackage{iopams}
\usepackage{graphicx}
\usepackage{cite}

\renewcommand{\vec}[1]{\bi{#1}}

\begin{document}

\title{The hydrogen atom in an electric field:
       Closed-orbit theory with bifurcating orbits}
\author{T Bartsch, J Main and G Wunner}
\address{Institut f\"ur Theoretische Physik 1, Universit\"at Stuttgart,
         70550 Stuttgart, Germany}
\ead{bartsch@theo1.physik.uni-stuttgart.de}

\begin{abstract}
Closed-orbit theory provides a general approach to the semiclassical
description of photo-absorption spectra of arbitrary atoms in external
fields, the simplest of which is the hydrogen atom in an electric
field. Yet, despite its apparent simplicity, a semiclassical quantization
of this system by means of closed-orbit theory has not been achieved so
far. It is the aim of this paper to close that gap. We first present a
detailed analytic study of the closed classical orbits and their
bifurcations. We then derive a simple form of the uniform semiclassical
approximation for the bifurcations that is suitable for an inclusion into a
closed-orbit summation. By means of a generalized version of the
semiclassical quantization by harmonic inversion, we succeed in calculating
high-quality semiclassical spectra for the hydrogen atom in an electric
field.
\end{abstract}

\submitto{\jpb}
%\maketitle

\section{Introduction}

Because it is classically integrable, the hydrogen atom in a homogeneous
electric field can be regarded as the simplest and most fundamental atomic
system apart from the field-free hydrogen atom, and consequently the
semiclassical treatment of that system is of particular importance. Due to the
integrability, of course, the Einstein-Brillouin-Keller torus quantization
rules apply, and a semiclassical calculation of photo-absorption spectra
based on the existence of quantized classical tori was indeed carried out
by Kondratovich and Delos \cite{Kondratovich97,Kondratovich98}.

A more general approach to the semiclassical description of atomic spectra
is furnished by closed-orbit theory \cite{Du88,Bogomolny89}, which is
intended to be applicable to atomic systems exhibiting either regular,
chaotic or mixed classical behaviour. Therefore, it is of vital interest to
see how it can be applied to this apparently simple example.

Closed-orbit theory provides a semiclassical approximation to the quantum
response function
\begin{equation}
  \label{gDef}
  g(E)=-\frac 1{\pi}
       \sum_n \frac{|\left<i|D|n\right>|^2}{E-E_n+\rmi\epsilon} \;,
\end{equation}
which involves the energy eigenvalues $E_n$ and the dipole matrix elements
$\left<i|D|n\right>$ connecting the excited states $n$ to the inital state
$i$.
Semiclassically, the response function splits into a smooth part and an
oscillatory part of the form
\begin{equation}
  \label{resOscGen}
  g^{\rm osc}(E) =
    \sum_{\rm c.o.} {\cal A}_{\rm c.o.}(E)\, \rme^{\rmi S_{\rm c.o.}(E)} \;,
\end{equation}
where the sum extends over all classical closed orbits starting from the
nucleus and returning to it after having been deflected by the external
fields.  $S_{\rm c.o.}$ is the classical action of the closed orbit. The
precise form of the recurrence amplitude $\cal A_{\rm c.o.}$ depends on the
geometry of the external field configuration
\cite{Du88,Bogomolny89,Bartsch02,Bartsch03b}. In the presence of a single
external field, it reads
\begin{equation}
  \label{AAxial}
  {\cal A}_{\rm c.o.} = 4\pi
    \frac{{\cal Y}^\ast(\vartheta_f)\, 
    {\cal Y}(\vartheta_i)}{|m_{12}|}
    \,\rme^{\rmi(\pi/2) \,\mu}
\end{equation}
if the orbit is directed along the field axis and
\begin{equation}
  \label{AmpNonAx}
  {\cal A}_{\rm c.o.} = 2(2\pi)^{3/2}
    \frac{\sqrt{\sin\vartheta_i\sin\vartheta_f}}{\sqrt{|m_{12}|}}\,
    {\cal Y}^\ast(\vartheta_f) {\cal Y}(\vartheta_i)
    \exp\left(\rmi\frac{\pi}2\mu+\rmi\frac{\pi}4\right)
\end{equation}
if it is not. Here,~$\vartheta_i$ and~$\vartheta_f$ are the inital and
final angles of the trajectory with respect to the field axis, $\mu$ is its
Maslov index, and $m_{12}$ an element of the monodromy matrix. The function
${\cal Y}(\vartheta)$ encodes the relevant properties of the inital state
and the exciting photon.  The amplitudes are independent of whether the
classical dynamics is regular, chaotic, or mixed. This is in stark contrast
to semiclassical trace formulae \cite{Gutzwiller90}, which represent the
quantum density of states as a sum over classical periodic orbits formally
analogous to~(\ref{resOscGen}), but require different semiclassical
amplitudes in the cases of regular or chaotic dynamics. This property makes
closed-orbit theory an ideal starting point for a semiclassical
quantization of typical atomic systems in external fields.

Nevertheless, even for the simplest of such systems, the hydrogen atom in
an electric field, any attempt to semiclassically quantize based on
closed-orbit theory so far has been unsuccessful. The reason for this
failure is that, even though the closed orbits of the Stark system are easy
to describe and classify \cite{Gao94}, a multitude of bifurcations
exist. Bifurcations of closed orbits lead to a divergence of the
semiclassical recurrence amplitudes and thus spoil the semiclassical
spectrum. Uniform semiclassical approximations \cite{Gao97,Shaw98a,Shaw98b}
smooth these divergences, but they have so far been restricted to the
calculation of low-resolution spectra and have prevented the use of any
high-resolution semiclassical quantization scheme. Only recently
\cite{Bartsch02a} were we able to introduce a generalization of the
semiclassical quantization by harmonic inversion which is capable of coping
with uniform approximations.  We have chosen the hydrogen atom in an
electric field to illustrate our quantization scheme because its closed
orbits and their bifurcations can easily be understood and its simplicity
makes it a archetypal problem of closed-orbit theory. On the other hand,
bifurcations of closed or periodic orbits are a generic phenomenon of mixed
regular-chaotic systems, so that their treatment in a semiclassical
quantization scheme is an important task with wide-rangig applications. The
principal difficulties of this task are already present in the Stark
system, so that it furnishes an ideal testing ground for the novel
technique.

In \cite{Bartsch02a} we focused on the presentation of the quantization
scheme itself. The central theme of the present paper is the classical and
semiclassical dynamics of the hydrogen atom in an electric field. In
section~\ref{sec:StarkEq} the classical dynamics of the Stark system is
described and the equations of motion are solved explicitly. In
section~\ref{sec:StarkClosed}, the closed orbits and their bifurcations are
discussed. Semi-analytic formulae for all relevant orbital parameters are
derived which have not been given in the literature so far. In
section~\ref{sec:StarkUniform}, a uniform approximation for the
closed-orbit bifurcations in the Stark system is constructed. Although this
has been done before \cite{Gao97,Shaw98a,Shaw98b}, we derive our result in
a much simpler way as was done previously and bring it into a form that is
considerably easier to apply.  Section~\ref{sec:StarkScl} concludes by
presenting both low- and high-resolution semiclassical spectra. They prove
that uniform approximations can indeed be included into a high-resolution
semiclassical quantization. These results solve the long-standing problem
of a closed-orbit theory quantization of the Stark effect and serve as a
prototype example for the application of the novel method to other systems.

\section{The equations of motion}
\label{sec:StarkEq}

If we assume the electric field $\vec F$ to be directed along the $z$-axis,
the Hamiltonian describing the motion of the atomic electron reads, in
atomic units,
\begin{equation}
  \label{Ham}
  H = \frac{1}{2}\vec p^2 -\frac 1r + F z \;,
\end{equation}
where $r^2=x^2+y^2+z^2$. It can be simplified by means of its scaling
properties: If the coordinates and momenta are scaled according to
$\tilde\vec x=F^{1/2}\vec x$, $\tilde\vec p=F^{-1/4}\vec p$, the classical
dynamics is found not to depend on the energy $E$ and the electric field
strength $F$ separately, but only on the scaled energy $\tilde
E=F^{-1/2}E$. In this work, we will use scaled classical quantities
throughout in sections \ref{sec:StarkEq} and~\ref{sec:StarkClosed}, where
purely classical calculations are presented. For the sake of simplicity, we
will not mark them with a tilde. From section~\ref{sec:StarkUniform} on,
the notational distinction between scaled and unscaled quantities will be
taken up again.

As is well known \cite{Born25,LandauI}, the Hamilton-Jacobi equation
corresponding to (\ref{Ham}) separates in semiparabolic coordinates
$\rho_1,\rho_2,\varphi$, where
%begin{equation}
% \label{ParKoord}
% x=\rho_1\rho_2\cos\varphi\;,\qquad
% y=\rho_1\rho_2\sin\varphi\;,\qquad
% z=\frac 12\left(\rho_2^2-\rho_1^2\right) \;.
%end{equation}
\begin{equation}
  \label{ParKoord}
  \rho_1 = \sqrt{r-z} \;, \qquad \rho_2=\sqrt{r+z} \;,
\end{equation}
and $\varphi$ is the azimuth angle around the field axis. Notice that
$\rho_1$ and $\rho_2$ are defined by~(\ref{ParKoord}) up to a choice of
sign only.

The Hamilton-Jacobi approach leads most naturally to the computation of the
action-angle variables fundamental to the EBK quantization. For a
discussion of closed orbits, however, it is essential to follow the
temporal evolution of a trajectory. This can most conveniently be achieved
if the trajectories are parameterized by a parameter~$\tau$ related to the
time~$t$ by
\begin{equation}
  \label{tauDef}
  d\tau = 2r\,dt \;.
\end{equation}
In the following, primes will be used to denote differentiation with
respect to $\tau$.
The Hamiltonian describing the dynamics as a function of $\tau$ is given by
\begin{equation}
  \label{regHam}
  \fl\eqalign{
  {\cal H} &= 2r\,\left(H-E\right) \\
      &= \frac 12 \left(p_{\rho_1}^2+p_{\rho_2}^2 
                  + \left(\frac 1{\rho_1^2}+\frac 1{\rho_2^2}\right)
                         p_\varphi^2
                  - 2E\left(\rho_1^2+\rho_2^2\right)
                  -\rho_1^4+\rho_2^4-4\right) = 0 \;. }
\end{equation}
(See \cite{Dirac33} or \cite{Bartsch02} for a discussion of how to treat
the pseudotime transformation~(\ref{tauDef}) within the framework of
Hamiltonian dynamics.)

Separating the pseudotime-Hamiltonian~(\ref{regHam}) into
$\rho_1$-dependent and $\rho_2$-dependent parts and using Hamilton's
equation of motion $\rho_j'=\partial{\cal H}/\partial p_{\rho_j} =
p_{\rho_j}$, we obtain
\begin{equation}
  \label{rhoInt1}
  {\rho_j'}^2 = \pm \rho_j^4 + 2 E \rho_j^2 - \frac{p_\varphi^2}{\rho_j^2}+c_j
  \qquad\textrm{ for $j=1,2$}
\end{equation}
with separation constants $c_1$ and $c_2$ related by 
\begin{equation}
  \label{cConst}
   c_1 + c_2 = 4 \;.
\end{equation}
Equation~(\ref{rhoInt1}) can be integrated to yield
\begin{equation}
  \label{tauInts}
       \tau = \int \frac{d\rho_1}
             {\sqrt{\phantom{-}\rho_1^4 + 2E\rho_1^2 - 
		\frac{p_\varphi^2}{\rho_1^2}+c_1}}
             = \int \frac{d\rho_2}
               {\sqrt{-\rho_2^4+2E\rho_2^2 -\frac{p_\varphi^2}{\rho_2^2}+c_2}}
\end{equation}
which gives the parameter $\tau$ as an elliptic integral in the
coordinates $\rho_1$ and $\rho_2$.

In the case of closed orbits, $p_\varphi=0$. At the nucleus, we have
$\rho_j'^2=c_j$ by~(\ref{rhoInt1}). The initial conditions for an orbit
starting at the nucleus at time $\tau=0$ thus read
\begin{equation}
  \label{StarkInitial}
  \eqalign{
    \rho_1(0)=0\;, & \rho_2(0)=0\;, \\
    \rho_1'(0)=\sqrt{c_1}\;, \qquad & \rho_2'(0)=\sqrt{c_2} \;.
  }
\end{equation}
The constants $c_1$ and $c_2$ describe the distribution of energy between the
uphill and the downhill motion.

According to~(\ref{rhoInt1}), for real
orbits $\rho_j$ can only assume values which make
\begin{equation}
  \label{fDef}
  f_j(\rho_j) = \pm \rho_j^4 + 2 E \rho_j^2 + c_j > 0 \;.
\end{equation}
For the motion in the $\rho_j$-direction to be bounded, there must be
a real turning point, where $f_j(\rho_j)=0$, because otherwise
$\rho_j$ will keep increasing forever. From the factorized form
\begin{equation}
  \label{f1Fakt}
  f_1(\rho_1) = (\rho_1^2-\rho_{1+}^2)(\rho_1^2-\rho_{1-}^2) \;,
\end{equation}
with
\begin{equation}
  \label{rho1Def}
  \rho_{1\pm}^2 = -E \pm \sqrt{E^2-c_1} \;,
\end{equation}
it is obvious that for $E<0$ the zeros of $f_1(\rho_1)$ are real if
\begin{equation}
  \label{c1Cond}
  c_1 < E^2
\end{equation}
and complex otherwise. Therefore, a closed orbit can only exist
if~(\ref{c1Cond}) holds. For energies $E$ below the Stark saddle energy
$E_{\rm S} = -2$, this condition is satisfied for all real orbits because
$c_2=4-c_1>0$.  For $E>E_{\rm S}$, there must be a sufficiently strong
component of the motion in the uphill direction so that the energy of the
downhill motion does not suffice to cross the saddle.

For the $\rho_2$ motion we obtain similarly
\begin{equation}
  \label{f2Fakt}
  f_2(\rho_2) = -(\rho_2^2-\rho_{2+}^2)(\rho_2^2-\rho_{2-}^2) \;,
\end{equation}
with
\begin{equation}
  \label{rho2Def}
  \rho_{2\pm}^2 = E \pm \sqrt{E^2+c_2} \;.
\end{equation}
The coordinate $\rho_{2+}$ is real for any $c_2>0$. Therefore, the
$\rho_2$-motion is always bounded for real orbits.

With the help of the factorizations (\ref{f1Fakt}) and (\ref{f2Fakt}), the
elliptic integrals (\ref{tauInts}) can easily be reduced to Legendre's
standard integral of the first kind \cite{Abramowitz}
\begin{equation}
  \label{ellInts}
  \fl\tau = \frac{1}{\rho_+} \,
     {\cal F}\left.\left(\arcsin \left(\frac{\rho_+}{\sqrt{c_1}}\,\rho_1\right)
             \right|m_1\right)
          = \frac{1}{\rho_-} \,
     {\cal F}\left.\left(\arcsin \left(\frac{\rho_-}{\sqrt{c_2}}\,\rho_2\right)
             \right|m_2\right) \;.
\end{equation}
The parameters are
\begin{equation}
  \label{mDef}
    m_1 = \frac{\rho_{1-}^2}{\rho_{1+}^2} =
           \frac{c_1}{\rho_+^4}\;, \qquad
    m_2 = \frac{\rho_{2+}^2}{\rho_{2-}^2} =
           -\frac{c_2}{\rho_-^4}\;,
\end{equation}
and the abbreviations
\begin{equation}
  \label{rhopmDef}
  \rho_+=\rho_{1+}\;,\qquad \rho_-=\sqrt{-\rho_{2-}^2}
\end{equation}
were introduced.
Equation (\ref{ellInts}) can be solved for $\rho_1$ and $\rho_2$ in terms
of Jacobi's elliptic functions \cite{Abramowitz} to yield
\begin{equation}
  \label{EllFkts}
  \eqalign{
    \rho_1(\tau) &= \frac{\sqrt{c_1}}{\rho_+}
      \textrm{sn}\,\left.\left(\rho_+\tau\right|m_1\right) \;, \\
    \rho_2(\tau) &= \frac{\sqrt{c_2}}{\rho_-}
      \textrm{sn}\,\left.\left(\rho_-\tau\right|m_2\right) \;.
  }
\end{equation}
These results incorporate the initial conditions (\ref{StarkInitial}).

\section{Closed orbits}
\label{sec:StarkClosed}

Closed orbits are characterized by the condition that at a pseudotime
$\tau_0>0$ the electron returns to the nucleus, so that 
\begin{equation}
  \label{closed1}
  \rho_1(\tau_0)=\rho_2(\tau_0)=0 \;.
\end{equation}
In the simplest cases, either $\rho_1$ or $\rho_2$ vanishes
identically. The electron then moves along the electric field axis.

If $c_1=0$, the downhill coordinate $\rho_1$ is zero. The electron leaves
the nucleus in the uphill direction, i.e. in the direction of the electric
field, until it is turned around by the joint action of the Coulomb and
external fields and returns to the nucleus at a pseudotime
\begin{equation}
  \label{tau2Def}
  \tau_2=2\,\frac{{\cal K}(m_2)}{\rho_-} \;,
\end{equation}
where ${\cal K}(m)$ denotes the complete elliptic integral of the first
kind \cite{Abramowitz}. Thus, $\tau_2$ corresponds to a half period of
$\rho_2$ and, due to the freedom in the choice of sign, to a full
period of the uphill orbit in position space. The uphill orbit is repeated
periodically and closes again at $\tau_0=l\tau_1$ with integer $l$.

The second case of axial motion is obtained if $c_1=4$, which corresponds
to a downhill motion opposite to the direction of the electric field. The
$\rho_1$ motion closes at pseudotimes $\tau=k\tau_1$ with
\begin{equation}
  \label{tau1Def}
  \tau_1=2\,\frac{{\cal K}(m_1)}{\rho_{+}} \;,
\end{equation}
if the energy $E$ is less than the saddle point energy $E_{\rm S}=-2$. If
$E>E_{\rm S}$, the electron crosses the Stark saddle and the orbit does not
close.

In the case of a non-axial orbit, the uphill and downhill
motions must close at the same time
\begin{equation}
  \label{closed2}
  \tau_0=k \tau_1=l \tau_2 \;.
\end{equation}
If (\ref{closed2}) is satisfied for any given integer values of $k$ and
$l$, the orbit returns to the nucleus after $k$ half periods of $\rho_1$
and $l$ half periods of $\rho_2$, corresponding, in Cartesian coordinates,
to $k$ full periods in the downhill and $l$ full periods in the uphill
direction. We therefore refer to $k$ and $l$ as the downhill and uphill
repetition numbers, respectively, and identify a non-axial closed orbit by
its repetition numbers $(k,l)$.

If the scaled energy $E$ is fixed, the half-periods
$\tau_1$ and $\tau_2$ depend on the separation constant $c_1$. For given
$k$ and $l$, (\ref{closed2}) can therefore be read as an equation for
$c_1$, thus determining the initial conditions of a closed orbit. We are now
going to investigate the solution of this equation to determine the
conditions for a $(k,l)$ orbit to exist.

\begin{figure}
  \centerline{\includegraphics[]{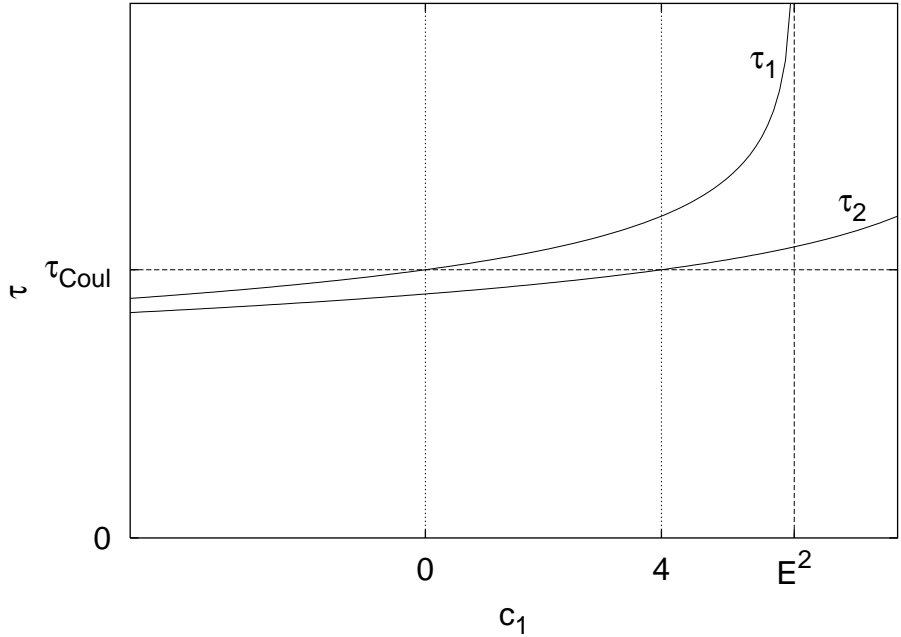}}
  \caption{Downhill period $\tau_1$ and uphill period $\tau_2$ as a
  function of the separation constant $c_1$.}
  \label{T12ofC1Fig}
\end{figure}

Figure \ref{T12ofC1Fig} shows the dependence of $\tau_1$ and $\tau_2$ on
$c_1$ for a fixed energy $ E$. For all $c_1$, the uphill period
$\tau_2$ is smaller than the downhill period $\tau_1$. Furthermore, as
$c_1\to-\infty$ we obtain from~(\ref{tau2Def}) and~(\ref{tau1Def}), to
leading order,
\begin{equation}
  \tau_1-\tau_2\approx\frac{2{\cal K}(-1)}{(-c_1)^{5/4}} \;,
\end{equation}
whereas
\begin{equation}
  \tau_1\approx\frac{2{\cal K}(-1)}{(-c_1)^{1/4}} \;,
\end{equation}
so that the ratio $\tau_1/\tau_2$ approaches 1 from above. On the other
hand, $\tau_1$ diverges at $c_1= E^2$, so that $\tau_1/\tau_2\to\infty$ as
$c_1\to E^2$. Thus, for any energy $E<0$ and positive integers $k$, $l$
with $l>k$, there is a unique solution $c_1$ to (\ref{closed2}). It is in
the range $-\infty<c_1<E^2$. There is no solution for $l\leq k$, since
$\tau_1$ is complex for $c_1> E^2$ whereas $\tau_2$ remains
real. Therefore, orbits $(k,l)$ with $l\leq k$ do not exist.

A real closed orbit must satisfy 
\begin{equation}
  \label{StarkConst}
  0\leq c_1\leq 4
\end{equation}
due to~(\ref{cConst}) and~(\ref{StarkInitial}). If (\ref{StarkConst}) is
not met, the initial conditions~(\ref{StarkInitial}) are complex and a
ghost orbit in the complexified phase space is obtained. More precisely, if
$c_1>4$, the initial velocity $\rho_2'(0)$ is imaginary, hence
$\rho_2(\tau)$ will be imaginary at all times. Similarly, $\rho_1(\tau)$
will be imaginary if $c_1<0$. In both cases, if the complex conjugate of
the ghost orbit is taken, one of the semiparabolic coordinates changes
sign. This change of sign does not alter the Cartesian coordinates, and
consequently all ghost orbits are invariant with respect to complex
conjugation.  By the same token, the starting angle of a ghost orbit is
defined up to complex conjugation only. We will always choose the imaginary
part of starting angles to be positive.  The properties of real and ghost
orbits are summarized in table~\ref{realGhostTable}.

\begin{table}
  \centerline{
  \begin{tabular}{|l||c|c|c|c|c|c|}
    \hline
    & $c_1$ & $c_2$ & $\rho_1$ & $\rho_2$ & $m_1$ & $m_2$\\ \hline
   pre-bifurcation ghost & $c_1>4$ & $c_2<0$ & $\in\mathbb{R}$ &
        $\in\rmi\mathbb{R}$ & $>0$ &
        $>0$ \\
   real orbit & $0\le c_1 \le 4$ & $0\le c_2\le 4$& $\in\mathbb{R}$ &
        $\in\mathbb{R}$ & $>0$ &
        $<0$ \\
   post-bifurcation ghost & $c_1<0$ & $c_2>4$ & $\in\rmi\mathbb{R}$ & 
        $\in\mathbb{R}$ & $<0$ & $<0$ \\ \hline
  \end{tabular}
  }
  \caption{Comparison of parameter ranges for real and ghost orbits}
  \label{realGhostTable}
\end{table}

To investigate the bifurcations of the closed orbits, we will now
discuss the dependence of $c_1$ on $E$ for fixed repetition numbers
$k,l$. For any energy $E$ and $c_1=0$, the downhill period $\tau_1$ is
equal to the period
\begin{equation}
  \label{tCoulDef}
  \tau_{\rm Coul}=\frac{\pi}{\sqrt{-2 E}}
\end{equation}
of the pure Coulomb dynamics. For $c_1=4$, the uphill period $\tau_2$
equals $\tau_{\rm Coul}$. For any other value of $c_1$, both $\tau_1$ and
$\tau_2$ converge to $\tau_{\rm Coul}$ as $E\to-\infty$, whence in this
limit $\tau_1/\tau_2\to 1$. Thus, $\tau_1/\tau_2<l/k$ at $c_1=4$ if $E$ is
sufficiently low. The solution to~(\ref{closed2}) must then lie in the
interval $c_1\in[4,E^2]$, so that the $(k,l)$-orbit is a ghost.  As $E$
increases, $c_1$ decreases. At the critical energy $E_{\rm gen}$ where
$c_1=4$, a real $(k,l)$ orbit is generated. It bifurcates off the downhill
orbit that is located at $c_1=4$. For $ E> E_{\rm gen}$, $c_1$ decreases
further. When the  energy $ E_{\rm dest}$ where $c_1=0$
is reached, the $(k,l)$ orbit collides with the uphill orbit and becomes a
ghost again. As the singularity of $\tau_1$ approaches zero as $ E\nearrow
0$, there is an $ E_{\rm dest}<0$ for any $(k,l)$.

The bifurcation energies $ E_{\rm gen}$ and $ E_{\rm dest}$ can be
determined from~(\ref{closed2}) if $c_1=4$ is prescribed for the generation
of an orbit, or $c_1=0$ for its destruction. For these cases,
equation~(\ref{closed2}) simplifies to
\begin{equation}
  \label{genEq}
  \frac{l}{k} = \frac{2^{3/2}}{\pi\sqrt{1+\sqrt{1-\epsilon}}}
    \,\,{\cal K}\left(\frac{\epsilon}
                           {\left(1+\sqrt{1-\epsilon}\right)^2}\right)
%  \frac{k}{\sqrt{- E+\sqrt{ E^2-4}}}\,
%  {\cal K}\left(\frac{4}{\left(- E+\sqrt{ E^2-4}\right)^2}\right)=
%  \frac{l}{\sqrt{-2 E}}\,\frac{\pi}{2}
\end{equation}
in terms of the dimensionless variable $\epsilon=4/E^2=(E_{\rm S}/E)^2$
for the generation of the $(k,l)$ orbit and
\begin{equation}
  \label{destEq}
  \frac{k}{l} = \frac{2^{3/2}}{\pi\sqrt{1+\sqrt{1+\epsilon}}}
    \,\,{\cal K}\left(\frac{-\epsilon}
                           {\left(1+\sqrt{1+\epsilon}\right)^2}\right)
%  \frac{k}{\sqrt{-2 E}}\,\frac{\pi}{2} =
%  \frac{l}{\sqrt{- E+\sqrt{ E^2+4}}}\,
%  {\cal K}\left(-\frac{4}{\left(- E+\sqrt{ E^2+4}\right)^2}\right)
\end{equation}
for its destruction. These equations provide a simple and stable method to
determine the bifurcation energies. They can be expected to yield more
accurate results then the  numerically computed monodromy
matrix elements used by Gao and Delos \cite{Gao94}.

\begin{figure}
  \centerline{\includegraphics[width=\textwidth]{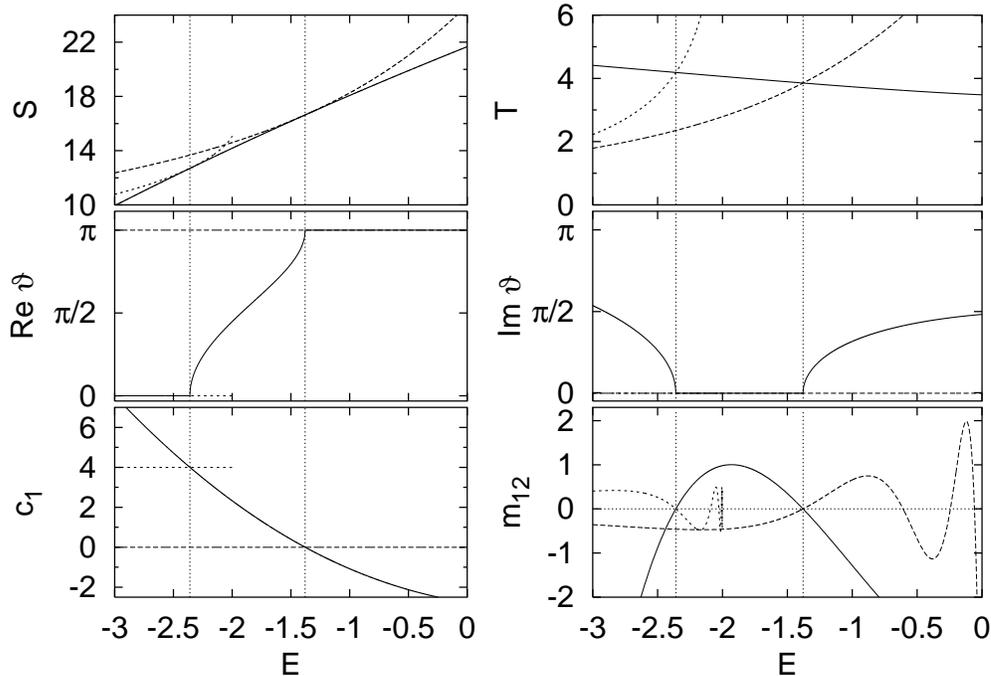}}
  \caption{Orbital parameters close to the bifurcations of the orbit
  $(4,5)$. Solid line: non-axial orbit $(4,5)$, long-dashed line: fifth
  uphill orbit, short-dashed line: fourth downhill orbit. Vertical lines
  indicate the bifurcation energies $ E_{\rm gen}=-2.3597$ and $ 
  E_{\rm dest}=-1.3790$, respectively}
  \label{StarkBifFig}
\end{figure}

The sequence of bifurcations described above is illustrated in figure
\ref{StarkBifFig}, where characteristic data of the orbit $(4,5)$, and the
downhill and uphill orbits it bifurcates from, is shown. The bifurcation
energies $ E_{\rm gen}=-2.3597$ and $ E_{\rm dest}=-1.3790$ are
characterized by the conditions $c_1=4$ and $c_1=0$, respectively. The
transition between real and ghost orbits can most clearly be seen from the
energy dependence of the starting angle $\vartheta_i$, which is real between
the bifurcation energies and acquires a non-zero imaginary part outside
this interval. In addition, at the bifurcation energies the actions and
orbital periods of the non-axial orbit coincide with those of the
appropriate axial orbits. Note that the action of the non-axial orbit is
always smaller than that of the axial orbits.

It is also apparent from figure \ref{StarkBifFig} that the monodromy matrix
element $m_{12}$ vanishes at the bifurcations. For the axial orbits,
further zeros of $m_{12}$ appear in figure \ref{StarkBifFig}. They
correspond to additional bifurcations these orbits undergo. As $k<l$ for a
non-axial orbit, the $l$th repetition of the uphill orbit undergoes $l-1$
bifurcations, where the orbits $(k,l)$ with $k=1,\dots,l-1$ are
destroyed. All four bifurcation energies of the fifth uphill orbit are
visible in figure~\ref{StarkBifFig}. A downhill orbit, on the contrary,
undergoes infinitely many bifurcations before it ceases to exist at
$E=-2$. Only the first members of this infinite cascade are resolved in
figure~\ref{StarkBifFig}.

The separation parameter $c_1$ for a closed orbit must be calculated
numerically from (\ref{closed2}). Once it is known, all orbital parameters
can be found analytically. The pertinent formulae will now be derived.

First of all, if we transform the orbits~(\ref{EllFkts}) to Cartesian
coordinates, we find that the starting and returning angle $\vartheta$ of a
closed orbit, measured with respect to the field axis, is related to the
separation constants by
\begin{equation}
  \label{thetaglg1}
  2\cos\frac{\vartheta}{2}=\sqrt{c_1}\;,\qquad
  2\sin\frac{\vartheta}{2}=\sqrt{c_2}\;,
\end{equation}
so that
\begin{equation}
  \label{thetaglg2}
  \vartheta=2\arccos\frac{\sqrt{c_1}}{2} \;.
\end{equation}
The angle $\vartheta$ determined from (\ref{thetaglg2}) is obviously real and
confined to the interval $0\leq \vartheta \leq\pi$ if $0\leq c_1\leq 4$, as
was to be expected for a real orbit. For a pre-bifurcation ghost orbit,
i.e. $c_1>4$, the equations (\ref{thetaglg1}) can be satisfied if
$\vartheta=\rmi\alpha$ is chosen purely imaginary with
\begin{equation}
  \label{alphapre}
  2\cosh\frac{\alpha}{2}=\sqrt{c_1}\;,\qquad
  2\rmi\sinh\frac{\alpha}{2}=\sqrt{c_2}\;.
\end{equation}
This choice of $\vartheta$, which is unique up to the addition of multiples
of $4\pi$, makes $\vartheta$ continuous at the bifurcation. Similarly, for
$c_1<0$, i.e. for a post-bifurcation ghost orbit, we set
$\vartheta=\pi+\rmi\alpha$ with
\begin{equation}
  \label{alphapost}
  2\rmi\sinh\frac{\alpha}{2}=\sqrt{c_1}\;,\qquad
  2\cosh\frac{\alpha}{2}=\sqrt{c_2}\;.
\end{equation}
Again, the real part makes $\vartheta$ continuous at the bifurcation.
In both cases we assume $\alpha>0$, which is possible because the orbit is
invariant under complex conjugation.

Using~$p_{\rho_j}=\rho_j'$, we obtain the
action integral
\begin{equation}
  \label{StarkAction1}
  \eqalign{
    S &= \int p_{\rho_1}\,d\rho_1 +
         \int p_{\rho_2}\,d\rho_2 \\
      &= 2k\int_0^{\sqrt{c_1}/\rho_+} \rho_1'\,d\rho_1 +
         2l\int_0^{\sqrt{c_2}/\rho_-} \rho_2'\,d\rho_2 \;.
  }
\end{equation}
The equations of motion (\ref{rhoInt1}) then yield
\begin{equation}
  \label{StarkAction2}
  \eqalign{
    S &= 2k\int_0^{\sqrt{c_1}/\rho_+} \sqrt{f_1(\rho_1)}\,d\rho_1 +
         2l\int_0^{\sqrt{c_2}/\rho_-} \sqrt{f_2(\rho_2)}\,d\rho_2 \\
     &= \frac{2}{3}\,k\rho_+^3\,I(m_1) -
        \frac{2}{3}\,l\rho_-^3\,I(m_2)
  }
\end{equation}
with
\begin{equation}
  I(m)=(m-1){\cal K}(m) + (m+1){\cal E}(m)
\end{equation}
and the complete elliptic integrals of the first and second kinds ${\cal
K}(m)$ and ${\cal E}(m)$ \cite{Abramowitz}. Note that $I(m)<0$ if $m<0$, so
that both terms in~(\ref{StarkAction2}) are positive for real orbits. For a
downhill orbit, the uphill repetition number $l$ is undefined. As, however, in
this case $m_2=0$ and $I(0)=0$, the uphill contribution drops out
of~(\ref{StarkAction2}) for a downhill orbit. Similarly, the downhill
contribution vanishes for uphill orbits.
The action $S$ given by~(\ref{StarkAction2}) is real for both real and ghost
orbits.

The physical-time period of a closed orbit is given by
\begin{equation}
  \label{StarkPeriod}
  \eqalign{
    T &= \int_0^{\tau_0} \rho_1^2\,d\rho_1 +
         \int_0^{\tau_0} \rho_2^2\,d\rho_2   \\
      &= 2k\rho_+\left({\cal K}(m_1)-{\cal E}(m_1)\right) -
         2l\rho_-\left({\cal K}(m_2)-{\cal E}(m_2)\right)\;.
  }
\end{equation}
As is the action, the period $T$ is always real.

The calculation of the monodromy matrix element $m_{12}$ is more
difficult. It proceeds from the $2\times 2$ Jacobian matrix 
\begin{equation}
  \label{JacobiDef}
  J = \frac{\partial\vec\rho(\tau_0)}{\partial\vec p(0)}
\end{equation}
that describes the change of the final position upon a variation of the
initial momentum. The initial and final momenta are given by
\begin{equation}
  \label{PiPf}
  \vec p(0)=\left(\begin{array}{c} \sqrt{c_1}\\ \sqrt{c_2}
                  \end{array}\right)\;, \qquad
  \vec p(\tau_0)=\left(\begin{array}{c} (-1)^k\sqrt{c_1} \\ 
                                 (-1)^l\sqrt{c_2}
                 \end{array}\right) \;,
\end{equation}
They have norm 2 by~(\ref{cConst}). As the monodromy matrix characterizes variations
perpendicular to the orbit, $m_{12}$ is obtained from $J$ by projecting
onto unit vectors perpendicular to the momenta~(\ref{PiPf}). Thus,
\begin{equation}
  \label{m12_1}
  m_{12}=\frac12\left(\begin{array}{cc} (-1)^l\sqrt{c_2}\;, & -(-1)^k\sqrt{c_1}
                \end{array}\right)
         \cdot J \cdot
         \frac12 \left(\begin{array}{c} \sqrt{c_2} \\ -\sqrt{c_1}
                 \end{array}\right)
  \;.
\end{equation}
Since the dynamics is separable, the matrix $J$ is diagonal, whence
\begin{equation}
  \label{m12_2}
  m_{12}=(-1)^l \frac{c_2}4\,
   	  \left.\frac{\partial \rho_1(\tau)}{\partial p_1(0)}\right|_
                {\tau=\tau_0}
        +(-1)^k \frac{c_1}4\,
	  \left.\frac{\partial \rho_2(\tau)}{\partial p_2(0)}\right|_
                {\tau=\tau_0}
  \;.
\end{equation}

For the remaining derivatives we find from (\ref{EllFkts})
\begin{equation}
  \label{m12_deriv}
  \fl
  \frac{\partial\rho_1(\tau)}{\partial p_1(0)} =
    \left(\frac{\partial}{\partial p_1(0)}\,\frac{\sqrt{c_1}}{\rho_+}\right)
      \textrm{sn}\,\left(\left.\rho_+\tau\right|m_1\right) +
    \frac{\sqrt{c_1}}{\rho_+}\,\frac{\partial}{\partial p_1(0)}\,
      \textrm{sn}\,\left(\left.\rho_+\tau\right|m_1\right) \;.
\end{equation}
A similar expression holds for $\rho_2$. If a non-axial orbit $(k,l)$ is
considered, the first term in (\ref{m12_deriv}) vanishes at $\tau=\tau_0$
by virtue of the
resonance condition (\ref{closed2}). The second term can be evaluated in an
elementary, but lengthy calculation to yield
\begin{equation}
  \label{m12}
  \eqalign{
    m_{12} =&\phantom{+}\;\,
              (-1)^{k+l}\,\frac{km_1c_2}{4\rho_+\sqrt{ E^2-c_1}}\,
              \left(2 E d(m_1)-\rho_+^2{\cal K}(m_1)\right) \\
           &+ (-1)^{k+l}\,\frac{lm_2c_1}{4\rho_-\sqrt{ E^2+c_2}}\,
              \left(2 E d(m_2)-\rho_-^2{\cal K}(m_2)\right)
  }
\end{equation}
with
\begin{equation}
  d(m) = \frac{{\cal E}(m)}{m(1-m)} - \frac{{\cal K}(m)}{m}\;.
\end{equation}

From (\ref{m12}) it can be verified that, up to a choice of sign, the
matrix element $m_{12}$ for the orbit $(nk,nl)$, which is the $n$th
repetition of $(k,l)$, equals $n$ times the matrix element for the orbit
$(k,l)$, as has been shown previously by Gao and Delos \cite{Gao94} by an
abstract argument using the neutral stability of the orbits. It is also
clear from (\ref{m12}) that $m_{12}$ vanishes when the orbit undergoes a
bifurcation, because $m_1\to 0$ as $c_1\to 0$, and ${\cal K}(m_1)$
and $d(m_1)$ both approach finite limits.

For an uphill orbit the second term in (\ref{m12_deriv})
vanishes because $c_1=0$, whereas the first term is nonzero in general because
the axial orbits do not obey a resonance condition akin
to~(\ref{closed2}). The derivative reads
\begin{equation}
  \label{deriv_uphill}
  \left.\frac{\partial \rho_1(\tau)}{\partial p_1(0)}\right|_{c_1=0} =
  \frac{\sin(\sqrt{-2 E}\,\tau)}{\sqrt{-2 E}} \;,
\end{equation}
so that
\begin{equation}
  \label{m12_uphill}
  m_{12} = (-1)^l\,
    \frac{\sin(\sqrt{-2 E}\,\tau_0)}{\sqrt{-2 E}} \;.
\end{equation}
Similarly, the matrix element for a downhill orbit is
\begin{equation}
  \label{m12_downhill}
  m_{12} = (-1)^k\,
    \frac{\sin(\sqrt{-2 E}\,\tau_0)}{\sqrt{-2 E}} \;.
\end{equation}

Finally, the Maslov indices of the closed orbits need to be determined. For
the uphill orbits, we proceed as follows: First, there is a caustic
whenever the orbit reaches either the nucleus or the turning point,
totalling to $2l-1$ caustics. Second, the Maslov index increases by two
(corresponding to two independent directions transverse to the orbit)
whenever the orbit is intersected by neighbouring trajectories, i.e. when
(\ref{deriv_uphill}) vanishes. This occurs, by~(\ref{deriv_uphill}), after
a pseudotime $\tau_{\rm Coul}$, so that the intersections contribute
$2[\tau_0/\tau_{\rm Coul}]$ to the Maslov index. $[x]$ denotes the integer
part of $x$. Third, if the semiclassical amplitudes are written according
to the conventions of \cite{Bartsch02,Bartsch03b}, the Maslov index must be
increased by one, so that for an uphill orbit it finally reads
\begin{equation}
  \label{uphillMaslov}
  \mu=2\left(l+\left[\frac{l\tau_2}{\tau_{\rm Coul}}\right]\right) \;.
\end{equation}
Using the same reasoning, we find
\begin{equation}
  \label{downhillMaslov}
  \mu=2\left(k+\left[\frac{k\tau_1}{\tau_{\rm Coul}}\right]\right) \;,
\end{equation}
for a downhill orbit. As the downhill period $\tau_1$ is always larger than
$\tau_{\rm Coul}$, the downhill Maslov index (\ref{downhillMaslov}) is
equal to $4k$ for sufficiently low $E$.

\begin{figure}
  \centerline{\includegraphics[]{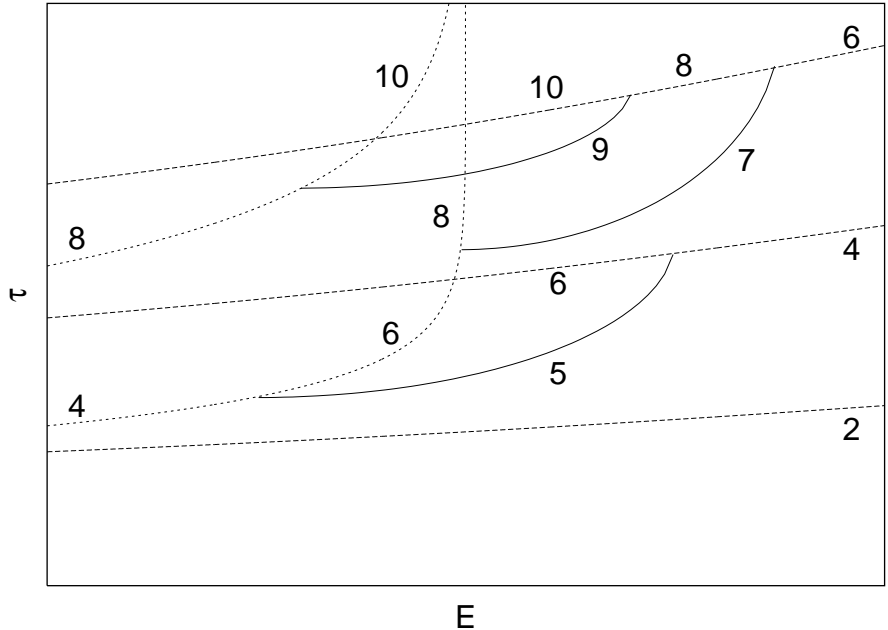}}
  \caption{Periods of the shortest closed orbits, labelled by  their Maslov
  indices, as a function of energy
  (schematic). Solid lines: non-axial orbits, short-dashed lines: downhill
  orbits, long-dashed lines: uphill orbits.}
  \label{MaslovFigure}
\end{figure}

The Maslov index of a non-axial orbit involves the number of zeros of
$m_{12}$ encountered along the orbit, which is hard to find
from~(\ref{EllFkts}). Instead, we exploit the observation that the change
of Maslov indices in a bifurcation can be determined from the normal form
describing the bifurcation. A normal form suitable for the bifurcations in
the Stark effect will be given in section~\ref{sec:StarkUniform}. It
predicts that in a bifurcation the Maslov index of the axial orbit
increases by two, which is consistent with~(\ref{uphillMaslov})
and~(\ref{downhillMaslov}), and the Maslov index of the non-axial orbit
takes the intermediate value. A schematic drawing of the shortest orbits
and their bifurcations in shown in figure \ref{MaslovFigure}. It implies
that the Maslov index of a non-axial orbit is
\begin{equation}
  \label{nonaxMaslov}
  \mu=2(k+l)-1 \;.
\end{equation}
Since along the orbit there are 
$k-1$ zeros of $\rho_1$ and $l-1$ zeros of $\rho_2$, each corresponding to
an intersection with the electric field axis and thus contributing~1 to the
Maslov index, there must be $k+l-1$ zeros of $m_{12}$. This result can be
confirmed numerically.

For ghost orbits, Maslov indices cannot be determined by counting caustics.
The correct phase of the ghost orbit contribution is fixed by demanding
that the semiclassical amplitude be continuous across the bifurcations, as
required by
the uniform approximation to be developed below.

\section{The uniform approximation}
\label{sec:StarkUniform}

In the following sections we present the semiclassical treatment of the
Stark effect. From now on, scaled quantities will again be marked with a
tilde.

It has become clear in section~\ref{sec:StarkClosed} that the apparently
simple dynamics of the hydrogen atom in an electric field contains numerous
bifurcations of closed orbits. They lead to zeros of the monodromy matrix
elements $m_{12}$ of the bifurcating orbits and thus to the divergence of
the pertinent closed-orbit amplitudes~(\ref{AmpNonAx}), so that they spoil the
semiclassical spectrum (see section~\ref{sec:StarkScl}).
To overcome this problem, a uniform semiclassical approximation giving the
collective contribution of the bifurcating orbits must be derived. For the
bifurcations occuring in the hydrogen atom in an electric field, this was
done by Gao and Delos \cite{Gao97}. These authors constructed a semiclassical
wave function near a bifurcation and essentially redid the derivation of
closed-orbit theory using that wave function. Their uniform approximation
was extended by Shaw and Robicheaux \cite{Shaw98a,Shaw98b} to include ghost
orbits.

In the form given in \cite{Gao97,Shaw98a,Shaw98b}, the uniform
approximation cannot be calculated from the actions and recurrence
amplitudes characterizing the bifurcating orbits, but requires the
determination of additional parameters. It is therefore hard to apply if
many closed-orbit bifurcations are to be dealt with. In this section we
will present a novel approach to uniformizing the bifurcations in the Stark
system that is not only simpler than previous derivations, but also yields
the uniform approximation in a more convenient form. It will be shown that
only the actions recurrence strengths of the isolated orbits are needed to
compute a uniform approximation. This makes the uniform approximation as
easy to use as the simple isolated-orbits formula.

The crucial step in the construction of a uniform approximation is the
choice of a suitable ansatz function whose stationary points describe the
classical closed orbits. In the Stark system, the structure of the
bifurcations is determined by the rotational symmetry of the
system. Non-axial orbits occur in one-parameter families generated by the
rotation of any particular member around the electric field axis, whereas
the axial orbit is isolated at all energies except at a bifurcation. The
normal form must share these symmetry properties.

We choose the normal form $\Phi$ to be defined on a plane with Cartesian
coordinates $(x,y)$ or polar coordinates
\begin{equation}
  \label{polarDef}
  x=r\,\cos\varphi\;, \qquad y=r\,\sin\varphi \;.
\end{equation}
Due to the rotational symmetry, $\Phi$ must be a function of the radial
coordinate~$r$ only, independent of the angle $\varphi$. To be a smooth
function of $x$ and $y$, it must be a function of $r^2$. Any such function
has a stationary point at the origin $x=y=0$ that corresponds to the
axial orbit because it is rotation invariant. Any stationary point at a
non-zero radial coordinate $r_{\rm c}$ occurs on a ring $r=r_{\rm c}$ and
describes a family of non-axial orbits. A normal form suitable to describe
the bifurcations occurring in the Stark system must therefore have
stationary points on a ring with radius $r_{\rm c}$ that shrinks to zero
as the bifurcation is approached.

These requirements are fulfilled by the normal form
\begin{equation}
  \label{StarkNF}
  \Phi_a(x,y)=\case14\, r^4 - \case12 a r^2
\end{equation}
with a real parameter $a$.
Apart from the trivial stationary point at the origin, (\ref{StarkNF}) has
a ring of stationary points at $r=\sqrt{a}$, which is real if $a>0$ and
imaginary if $a<0$. Thus, (\ref{StarkNF}) describes a family of non-axial
real orbits present for $a>0$ which then contracts onto an axial orbit and
becomes a family of ghost orbits.

For the uniform approximation we make the ansatz
\begin{equation}
  \label{StarkUnif}
  \Psi(E) = I(a)\,\rme^{\rmi S_{\rm ax}(E)} \;,
\end{equation}
where $S_{\rm ax}(E)$ denotes the action of the axial orbit and
\begin{equation}
  \label{StarkInt}
  I(a) = \int dx\,dy\, p(x^2+y^2)\,\rme^{\rmi\Phi_a(x,y)} \;,
\end{equation}
with an arbitrary smooth rotationally symmetric function $p(r^2)$.
The parameter $a$ and the function $p(r^2)$ must now be determined such
that, when the distance from the bifurcation gets large, the uniform
approximation reproduces the isolated-orbits formula. The asymptotic
behaviour of~(\ref{StarkUnif}) can be found if~(\ref{StarkInt}) is
evaluated in the stationary-phase approximation. The contribution of the
stationary point at the origin is straight-forward to evaluate. It yields
\begin{equation}
  \label{StarkOrigin}
  \left.I(a)\right|_{r=0}  = \frac{2\pi\rmi}{|a|}\,p(0)\,
		\rme^{-\rmi\case\pi 2\nu_0}
	= -\frac{2\pi\rmi}{a}\,p(0) \;,
\end{equation}
where
\begin{equation}
  \nu_0=\cases{
	  0 & : \textrm{$a<0$} \\
 	  2 & : \textrm{$a>0$}
 	}
\end{equation}
is the number of negative eigenvalues of the Hessian matrix. From this
number it is apparent that the Maslov index of the axial orbit increases by
2 in the bifurcation.

The contribution of the non-axial orbits is present if $a>0$. Due to the
rotational symmetry, a straight-forward stationary-phase approximation to
this contribution fails because the stationary points are not isolated.
 In polar coordinates, the integration over the angle $\varphi$
yields the constant $2\pi$, and the stationary-phase approximation can be
applied to the remaining integral over $r$. This yields
\begin{equation}
  \label{StarkNonax}
  \left.I(a)\right|_{\rm ring} = 
	2\pi\sqrt{\pi\rmi}\,p(a)\,\rme^{-\rmi a^2/4} \;.
\end{equation}
Note that the contribution of the non-axial orbit remains finite as $a\to0$.

Thus, the uniform approximation~(\ref{StarkUnif}) asymptotically equals
\begin{equation}
  \Psi(E) \approx 
     -\frac{2\pi\rmi}{a}\,p(0)\,\rme^{\rmi S_{\rm ax}(E)}
     +2\pi\sqrt{\pi\rmi}\,p(a)\,\rme^{\rmi (S_{\rm ax}(E)-a^2/4)} \;.
\end{equation}
This result should reproduce the isolated-orbits formula
\begin{equation}
  {\cal A}_{\rm ax}(E)\, \rme^{\rmi S_{\rm ax}(E)} +
  {\cal A}_{\rm non}(E)\, \rme^{\rmi S_{\rm non}(E)}
\end{equation}
in terms of the actions and recurrence amplitudes of the axial and
non-axial orbits. This can be achieved if we demand
\begin{eqnarray}
  \label{StarkActCond}
  S_{\rm non} = S_{\rm ax}-a^2/4 \;,\\
  \label{StarkAmpCond}
  {\cal A}_{\rm ax} = -\frac{2\pi\rmi}{a}\,p(0) \;, \qquad
  {\cal A}_{\rm non} = 2\pi\sqrt{\pi\rmi}\,p(a) \;.
\end{eqnarray}
The normal form parameter
\begin{equation}
  \label{StarkADef}
  a = \pm 2\sqrt{S_{\rm ax}-S_{\rm non}}
\end{equation}
can be determined from~(\ref{StarkActCond}). The sign of $a$ has to be
chosen according to whether the non-axial orbits are real or complex.

In the limit $a\to 0$, i.e.~close to the bifurcation, $p(a)\approx
p(0)$. Equation~(\ref{StarkAmpCond}) thus imposes the constraint
\begin{equation}
  \label{StarkAmpConst}
  \frac{1+\rmi}{\sqrt{2\pi}}\,{\cal A}_{\rm non} = - a {\cal A}_{\rm ax}
\end{equation}
on the semiclassical amplitudes. Note that both sides of this equation
are finite as $a\to0$. In particular, ${\cal A}_{\rm non}$ is continuous at
$a=0$. This condition was used in section~\ref{sec:StarkClosed} to fix the
Maslov indices for the ghost orbits.  The actual semiclassical amplitudes
satisfy (\ref{StarkAmpConst}) close to the bifurcation, but the agreement
is satisfactory in the immediate neighbourhood of the bifurcation only.

For $p(r^2)$ in equation~(\ref{StarkInt}) we choose a first-order
polynomial
\begin{equation}
  \label{StarkPDef}
  p(r^2) = p_0 + p_1 (r^2-a)
\end{equation}
for the amplitude function. The coefficients $p_0$ and $p_1$ are chosen to
satisfy (\ref{StarkAmpCond}):
\begin{equation}
  \label{StarkCoefDef}
  p_0=\frac{{\cal A}_{\rm non}}{\sqrt{2\pi}\,\pi(1+\rmi)} \;, \qquad
  p_1=\frac{1}{2\pi\rmi a}
   \left(a {\cal A}_{\rm ax}+\frac{1+\rmi}{\sqrt{2\pi}}\,{\cal A}_{\rm non}
   \right) \;.
\end{equation}
By (\ref{StarkAmpConst}), $p_1$ remains finite at the bifurcation.
Thus, the uniform approximation (\ref{StarkUnif}) assumes the form
\begin{equation}
  \label{StarkUnifFinal}
  \Psi(E) = 
    \left[\frac{{\cal A}_{\rm non}}{(1+\rmi)}\,I_0
         +\frac{1}{a}\left(a {\cal A}_{\rm ax}
                           +\frac{1+\rmi}{\sqrt{2\pi}}\,{\cal A}_{\rm non}
                     \right) I_1
    \right] \rme^{\rmi S_{\rm ax}}
\end{equation}
with the integrals
\begin{equation}
  \label{StarkIntDef}
  \fl
  I_0=\frac{1}{2^{1/2}\,\pi^{3/2}}\int dx\,dy\,\rme^{\rmi \Phi_a(x,y)} \;,
      \qquad
  I_1=\frac{1}{2\pi\rmi}\int dx\,dy\,(r^2-a)\,\rme^{\rmi \Phi_a(x,y)} \;.
\end{equation}
$I_0$ can be evaluated in terms of the Fresnel integrals \cite{Abramowitz}
\begin{equation}
  \label{FresnelDef}
  C(x)=\int_0^x \cos\left(\case\pi 2\,t^2\right) dt \;,\qquad
  S(x)=\int_0^x \sin\left(\case\pi 2\,t^2\right) dt
\end{equation}
to yield
\begin{equation}
  \label{StarkI0}
  I_0=\rme^{-\rmi a^2/4}
      \left[\frac{1+\rmi}{2}-C\left(-\frac{a}{\sqrt{2\pi}}\right)
                         -\rmi S\left(-\frac{a}{\sqrt{2\pi}}\right)
      \right] \;.
\end{equation}
$I_1$ can be reduced to
\begin{equation}
  \label{StarkI1}
  I_1=\frac{1}{2\rmi}\,\rme^{-\rmi a^2/4}
      \int_{-a}^{\infty}\,dv\,v\,\rme^{\rmi v^2/4}
\end{equation}
with $v=r^2-a$. The integral in (\ref{StarkI1}) diverges at $v=\infty$. It
can be regularized by adding a small exponential damping factor
$\rme^{-\varepsilon v^2}$ and letting $\varepsilon\to0$ at the end. This
procedure yields
\begin{equation}
  \label{StarkI1Final}
  I_1=1 \;.
\end{equation}
The regularization can be justified by noting that the stationary phase
ap\-proxi\-ma\-tion to (\ref{StarkIntDef}) also yields
$I_1=1$, so that (\ref{StarkI1Final}) has the required
asymptotic behaviour. 

\begin{figure}
  \centerline{\includegraphics[]{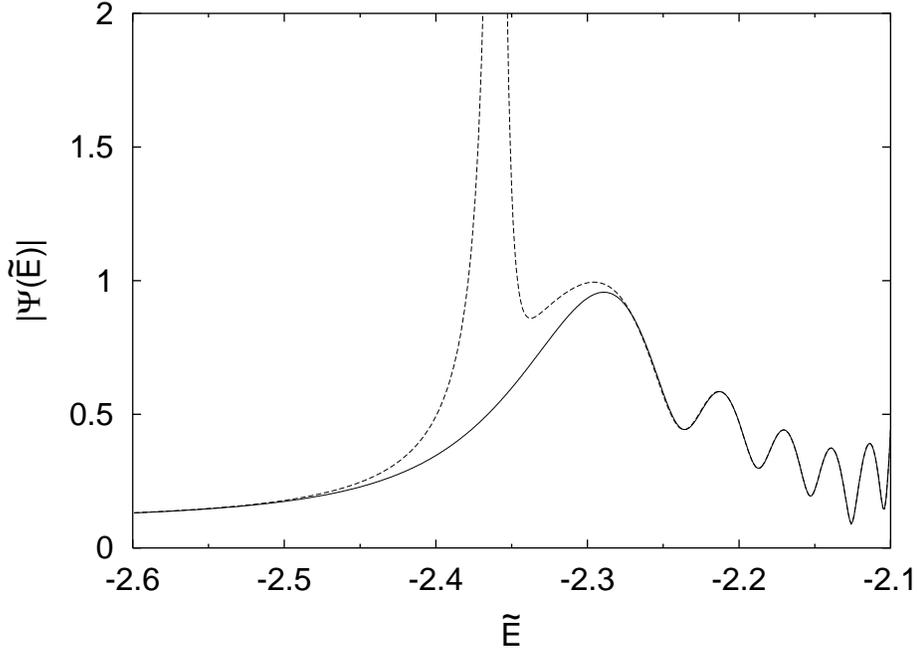}}
  \caption{Uniform approximation for the generation of the (4,5) non-axial
  	   orbit from the downhill orbit for the electric field strength
	   $F=10^{-8}$. Solid line: Uniform approximation
           (\ref{StarkUnifFinal}), dashed line: isolated-orbits
           approximation.}
  \label{Gen45Figure}
\end{figure}

The calculation of the uniform approximation (\ref{StarkUnif}) is thus
finished. As an example, the uniform approximation for the bifurcation of
the (4,5) non-axial orbit off the downhill orbit is compared to the simple
closed-orbit formula in figure~\ref{Gen45Figure}. Obviously, the uniform
approximation (\ref{StarkUnifFinal}) smooths the divergence of the
isolated-orbits approximation and, as desired, asymptotically reproduces
the simple approximation perfectly.

\section{Semiclassical spectra}
\label{sec:StarkScl}

A low-resolution semiclassical photo-absorption spectrum can be obtained
from the closed-orbit sum~(\ref{resOscGen}) by including orbits with an
orbital period up to a maximum $T_{\rm max}$ only. In order to resolve
individual energy levels, $T_{\rm max}$ must be larger than the Heisenberg
time $T_{\rm H}$. A rough estimate for $T_{\rm H}$ can be obtained from
perturbation theory. To first order in the electric field strength, the
energy splitting between two adjacent spectral lines with principal quantum
number $n$ is \cite{Born25} $\Delta E_{\rm p}=3nF$, so that the scaled
perturbative Heisenberg time is
\begin{equation}
  \label{THpert}
  \widetilde T_{\rm H, p} = \frac{2\pi F^{3/4}}{\Delta E_{\rm p}}
     = \frac{2\pi}{3}\sqrt{-2\tilde E} \;.
\end{equation}
This estimate is reasonable (although it may not be quantitatively
precise) as long as different $n$-manifolds do not overlap. If they do, the
mean level spacing is much smaller and the Heisenberg time therefore much
larger than given by (\ref{THpert}).

\begin{figure}
  \centerline{\includegraphics[]{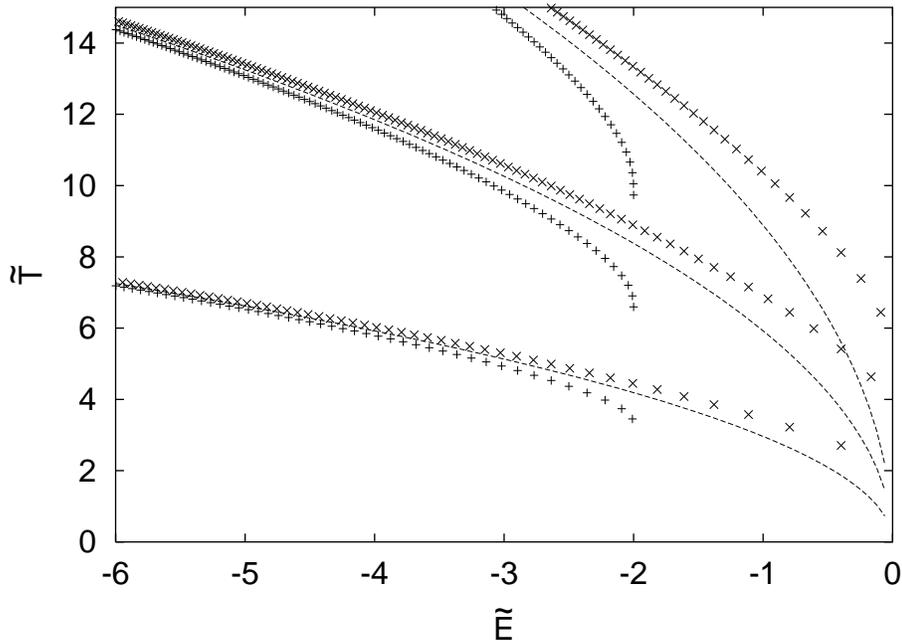}}
  \caption{Bifurcations of non-axial orbits off the downhill (+) and uphill
  ($\times$) orbits in a scaled period vs.~scaled energy plot.
  Bifurcation energies and scaled periods of the
  bifurcating orbits are indicated. Dashed lines: single, double and triple
  scaled
  perturbative Heisenberg time (\ref{THpert}).}
  \label{HeisenbergFig}
\end{figure}

Multiples of the scaled perturbative Heisenberg time (\ref{THpert}) are
plotted in figure \ref{HeisenbergFig} together with the scaled energies and
scaled periods of bifurcating orbits. For low scaled energies, the periods
of bifurcating orbits are well approximated by $\widetilde T_{\rm
H,p}$. Therefore, there is no parameter range where the closed-orbit sum
can be extended up to the Heisenberg time without involving bifurcations,
so that the uniformization of bifurcations must be an essential ingredient
to any closed-orbit theory quantization of the Stark effect.

\begin{figure}
  \centerline{\includegraphics[]{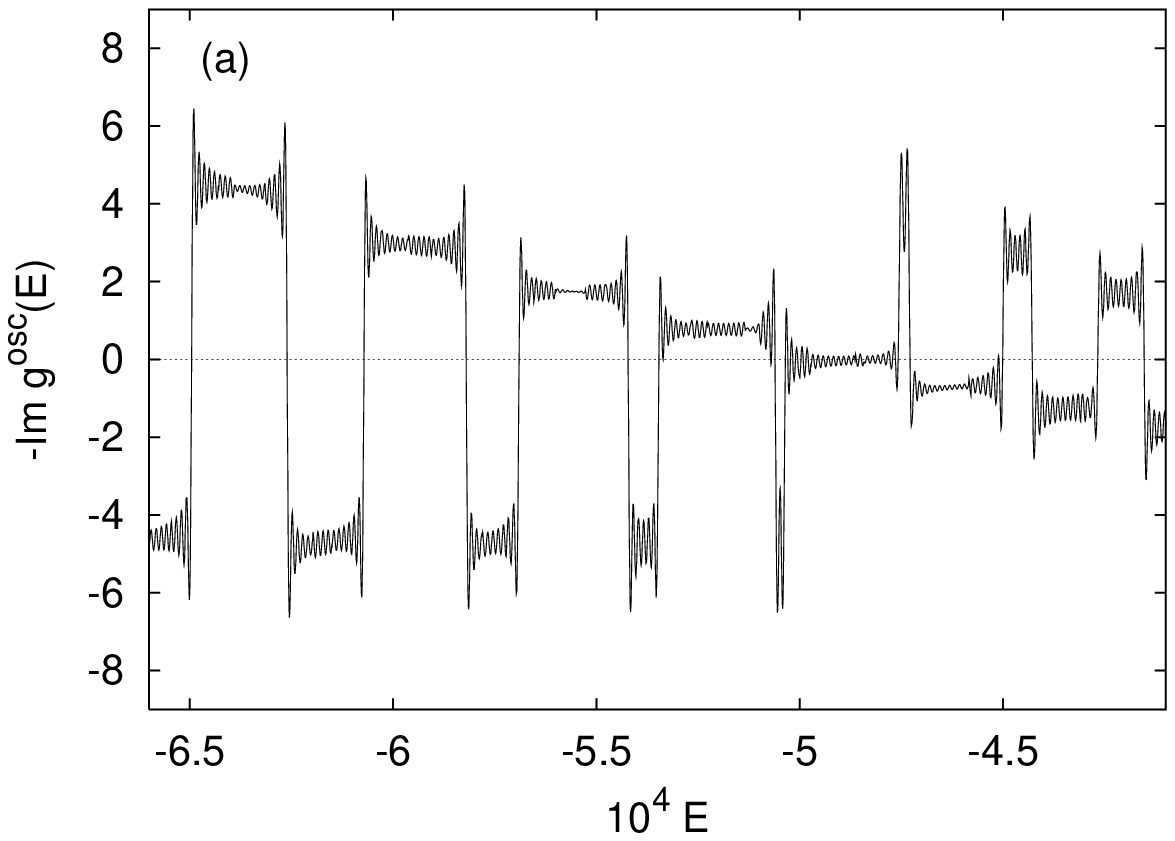}}
  \centerline{\includegraphics[]{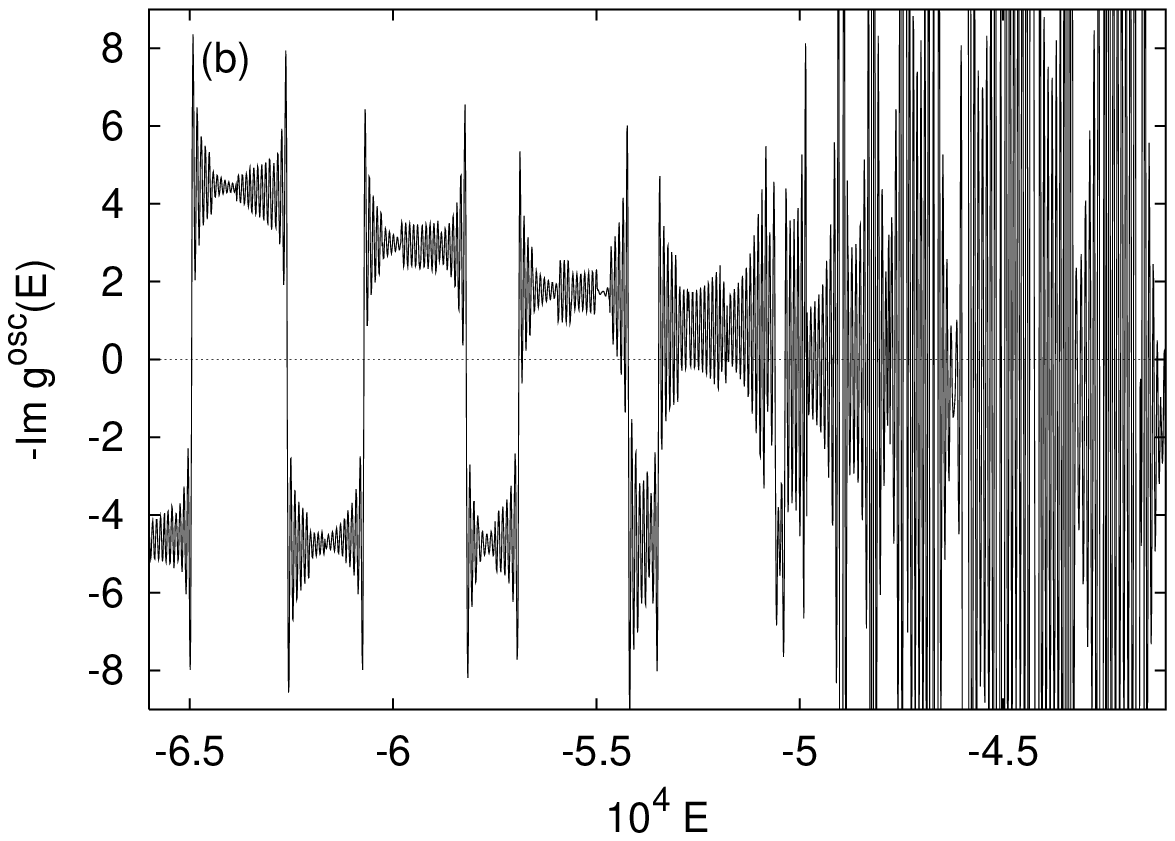}}
  \caption{Closed-orbit sum without uniform approximations as a function of
  energy
  at the electric field strength of $F=10^{-8}$
  with scaled cut-off time (a) $\widetilde T_{\rm max}=5$, 
  (b) $\widetilde T_{\rm
  max}=6.5$. At $E>-5\times 10^{-4}$, the latter is dominated by 
  bifurcation-induced divergences.}
  \label{StarkIsoFig}
\end{figure}

These findings can be confirmed numerically. Figure \ref{StarkIsoFig}
displays the oscillatory part of the photo-absorption cross section as
calculated from the closed-orbit sum. All spectra presented in this section
are non-scaled spectra calculated for the hydrogen atom in an electric
field $F=10^{-8}\hat{=} 51.4 \textrm{V/cm}$, corresponding to $w=100$ and
$\tilde E=10^4 E$, with the initial state $|1s\rangle$ and light linearly
polarized along the electric field axis. Small discontinuities are
introduced in the low-resolution spectra because a closed orbit abruptly
disappears from the truncated sum when its period increases beyond the
chosen cut-off time $\widetilde T_{\rm max}$. The discontinuities can be
avoided by choosing a smooth cut-off function to gradually switch off the
contribution of an orbit when its period approaches the cut-off time. The
changes in the spectra are hardly visible, and as excellent results can be
obtained with the simple hard cut-off, no smoothing will be used
henceforth.

For the spectrum in figure \ref{StarkIsoFig}(a), a scaled cut-off time of
$\widetilde T_{\rm max}=5$, well below the perturbative Heisenberg time,
was chosen. Consequently, one can distinguish groups of levels
characterized by a fixed principal quantum number $n$, but no individual
spectral lines can be resolved in the plot. The small oscillations are
an artefact of the closed-orbit sum and cannot be identified with
spectral lines. This is clear from the observations that the frequency of
the oscillation increases with increasing cut-off time and that virtually
the same oscillations are visible within an $n$-manifold and between
manifolds.

 At low energies, different $n$-manifolds do not overlap, so that spectral
regions with high oscillator strength density alternate with regions where
no spectral lines are present. (As the smooth part of the semiclassical
spectrum has been omitted, the bottom line of the semiclassical spectrum is
shifted to negative values of the oscillator strength density.) At
$E\approx -4.7\times 10^{-4}$, neighbouring $n$-manifolds start to
overlap. In the overlap regions the spectral density is considerably higher
than in regions of isolated $n$-manifolds.

To improve the resolution, the cut-off time must be
increased. Figure~\ref{StarkIsoFig}(b) displays the results for a scaled
cut-off time of $\widetilde T_{\rm max}=6.5$. At low energies, this is still
below the perturbative Heisenberg time, and no significant improvement of
the resolution can be found. At high energies, bifurcations start to occur,
and a dense sequence of bifurcation-induced divergences covers the
semiclassical signal. As the cut-off time is further increased,
bifurcations occur at ever lower energies and destroy ever larger parts of
the semiclassical spectrum. It is thus obvious that in its simple form the
closed-orbit sum is useless for a complete quantization.

\begin{figure}
  \centerline{\includegraphics[]{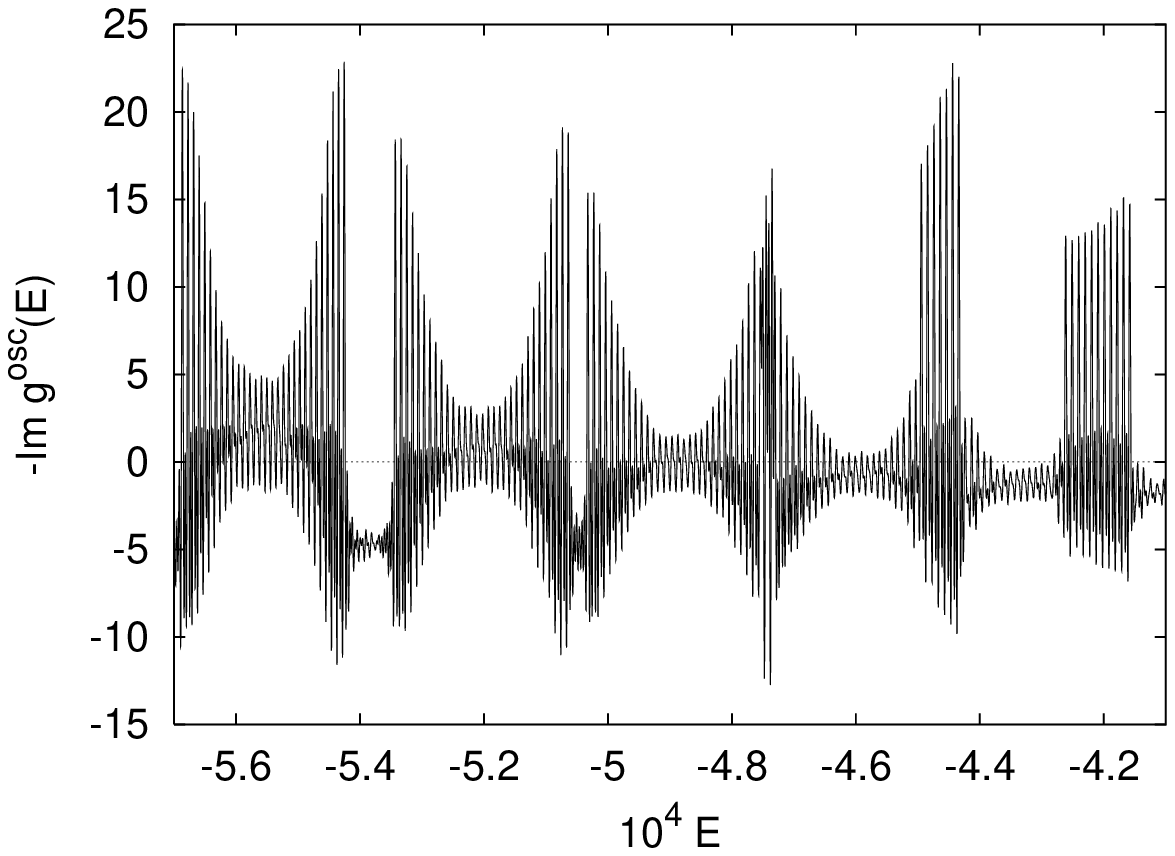}}
  \centerline{\includegraphics[]{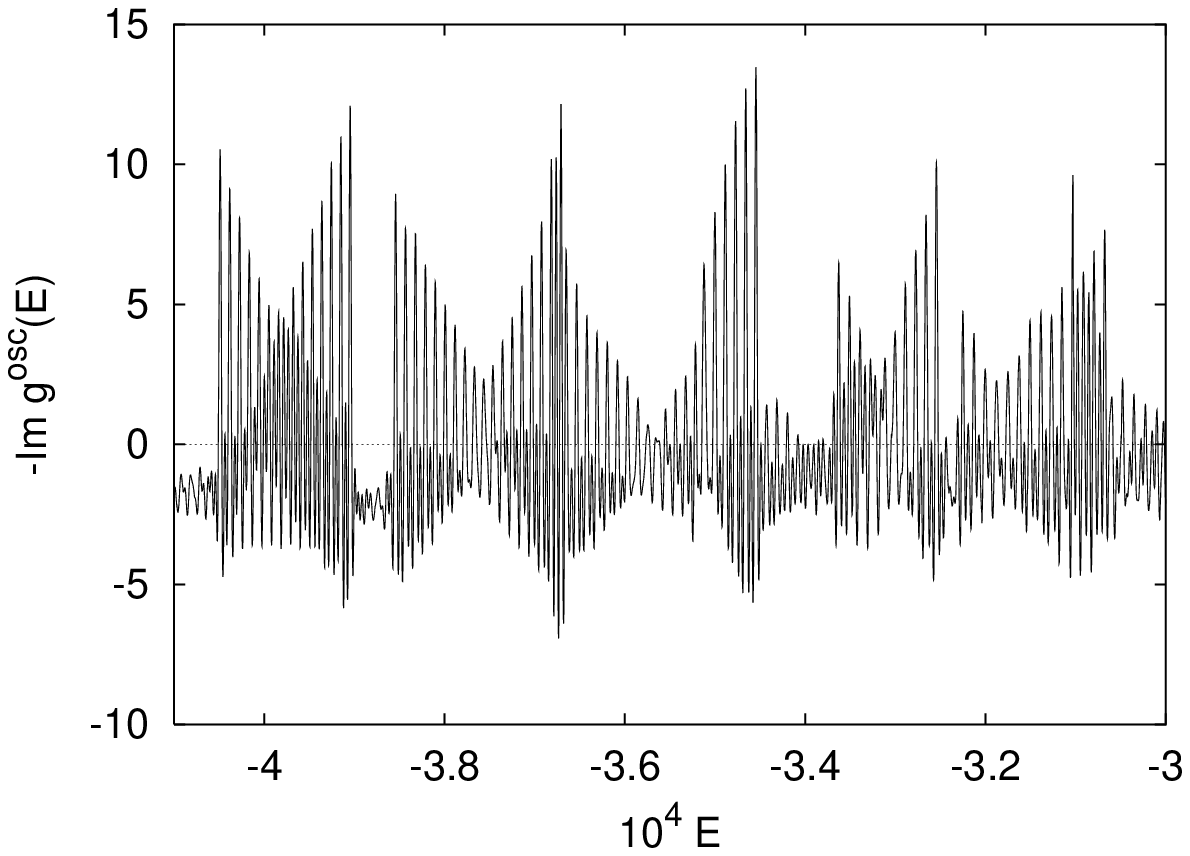}}
  \caption{Low-resolution semiclassical spectrum with scaled cut-off time
  $\widetilde T_{\rm max}=15$.}
  \label{StarkT15LoFig}
\end{figure}

If, on the contrary, the uniform approximations are included, the
closed-orbit sum can be extended to longer
orbits. Figure~\ref{StarkT15LoFig} shows the uniformized closed-orbit sum
for a scaled cut-off time $\widetilde T_{\rm max}=15$. As will also be done
in all subsequent semiclassical spectra, the uniform approximation was
applied for bifurcating orbits whose action difference is less than
$2\pi$. If for a given orbit several other orbits satisfy this requirement,
a more complicated uniform approximation describing several bifurcations
should be used (see the discussion at the end of section~\ref{sec:StarkScl}). 
For the time being, we resolve this conflict by calculating
the simple uniform approximation for the two orbits with the smallest
action difference and treating all other orbits as isolated.

\begin{figure}
  \centerline{\includegraphics[]{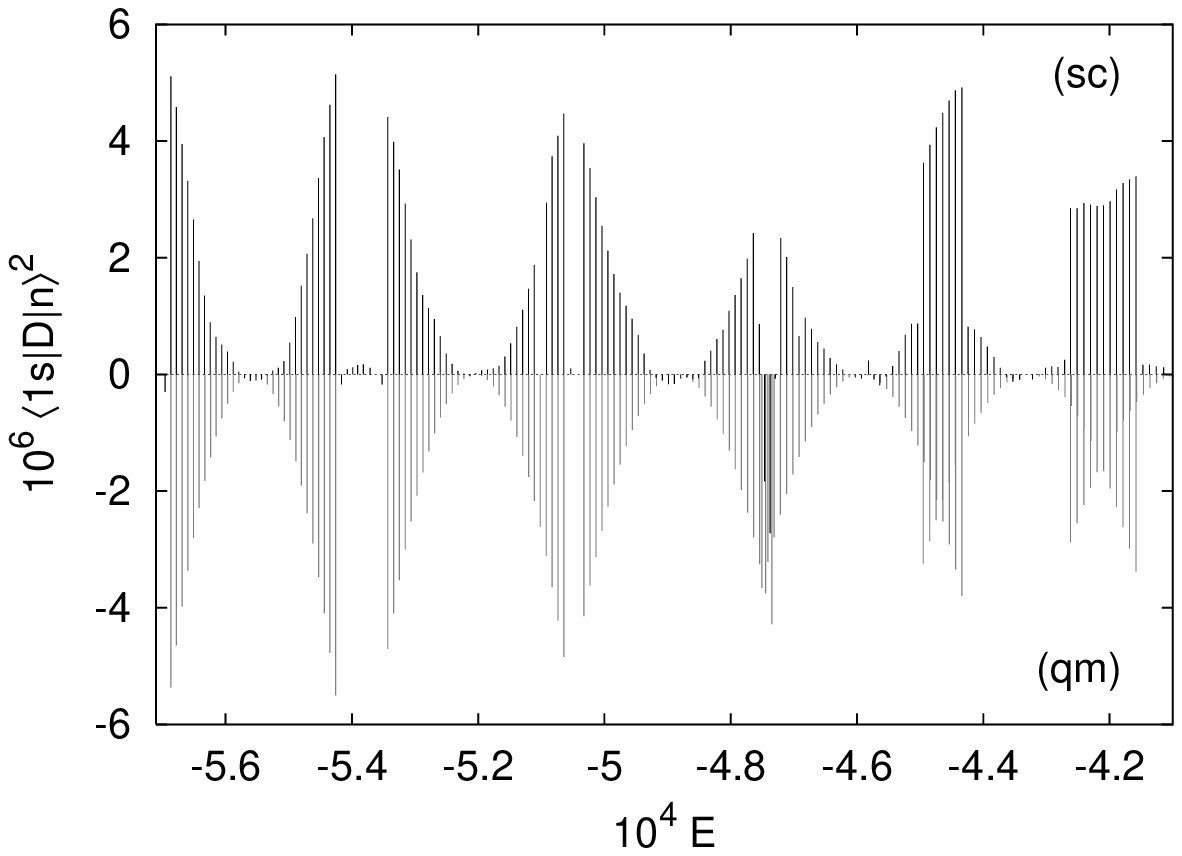}}
  \centerline{\includegraphics[]{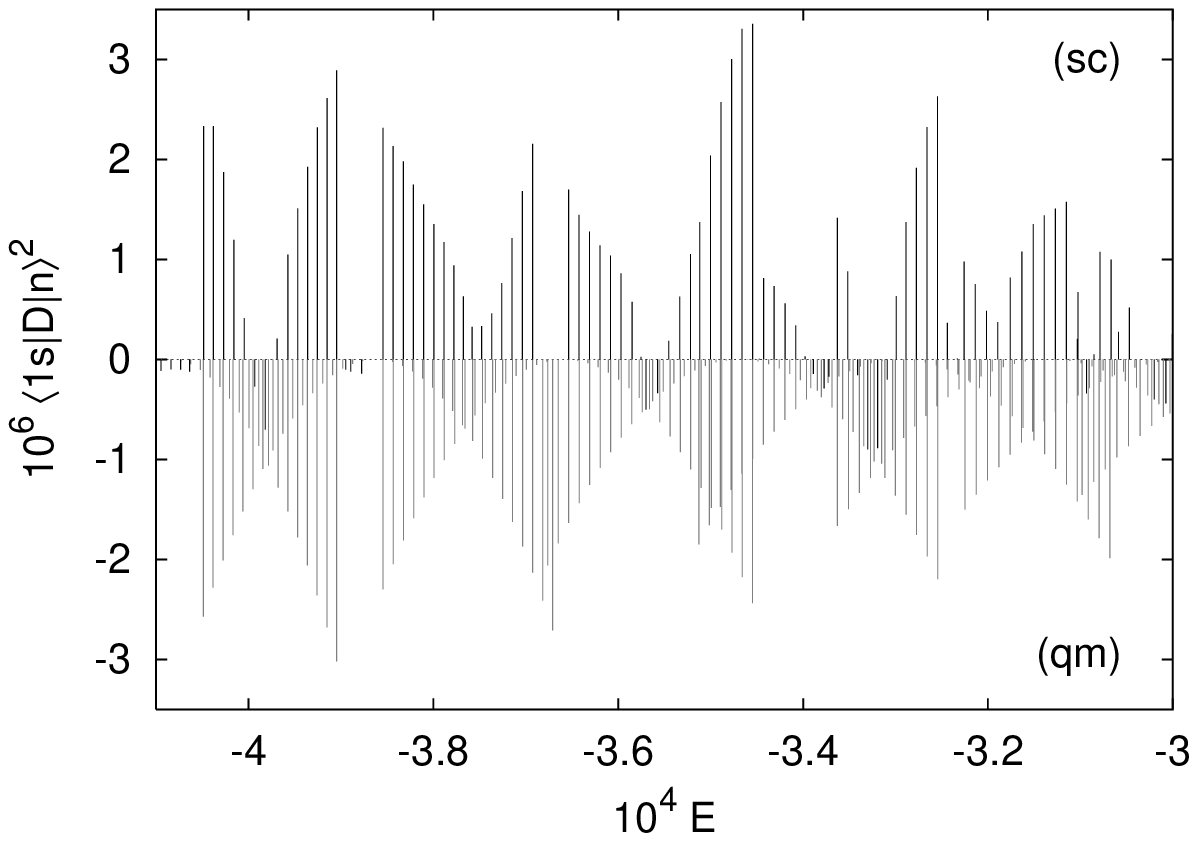}}
  \caption{High-resolution semiclassical (sc) and quantum (qm,
  inverted) photo-absorption spectrum. The scaled cut-off time for
  the semiclassical spectrum is $\widetilde T_{\rm max}=15$.}
  \label{StarkT15HiFig}
\end{figure}

Although individual spectral lines can be discerned in figure
\ref{StarkT15LoFig}, it would be hard to obtain precise values of the
energy levels and, in particular, of the associated dipole matrix elements
from this figure. A more reliable method of extracting the spectral
information from the semiclassical data is clearly desirable. To this end,
we will use the method of semiclassical quantization by harmonic inversion
\cite{Main97c,Main98b,Main99d}, which we recently generalized to be
applicable in connection with uniform approximations
\cite{Bartsch02a}. This method requires the cut-off time of the closed
orbit sum to be chosen larger than twice the Heisenberg time for this
purpose.  The perturbative Heisenberg time thus suggests that a scaled
cut-off time of $\widetilde T_{\rm max}=15$ will be sufficient as long as
different $n$-manifolds do not overlap. Figure~\ref{StarkT15HiFig} presents
the energy levels and transition matrix elements obtained from it by the
harmonic inversion procedure and compares them to the exact quantum
results.

\begin{figure}
  \centerline{\includegraphics[]{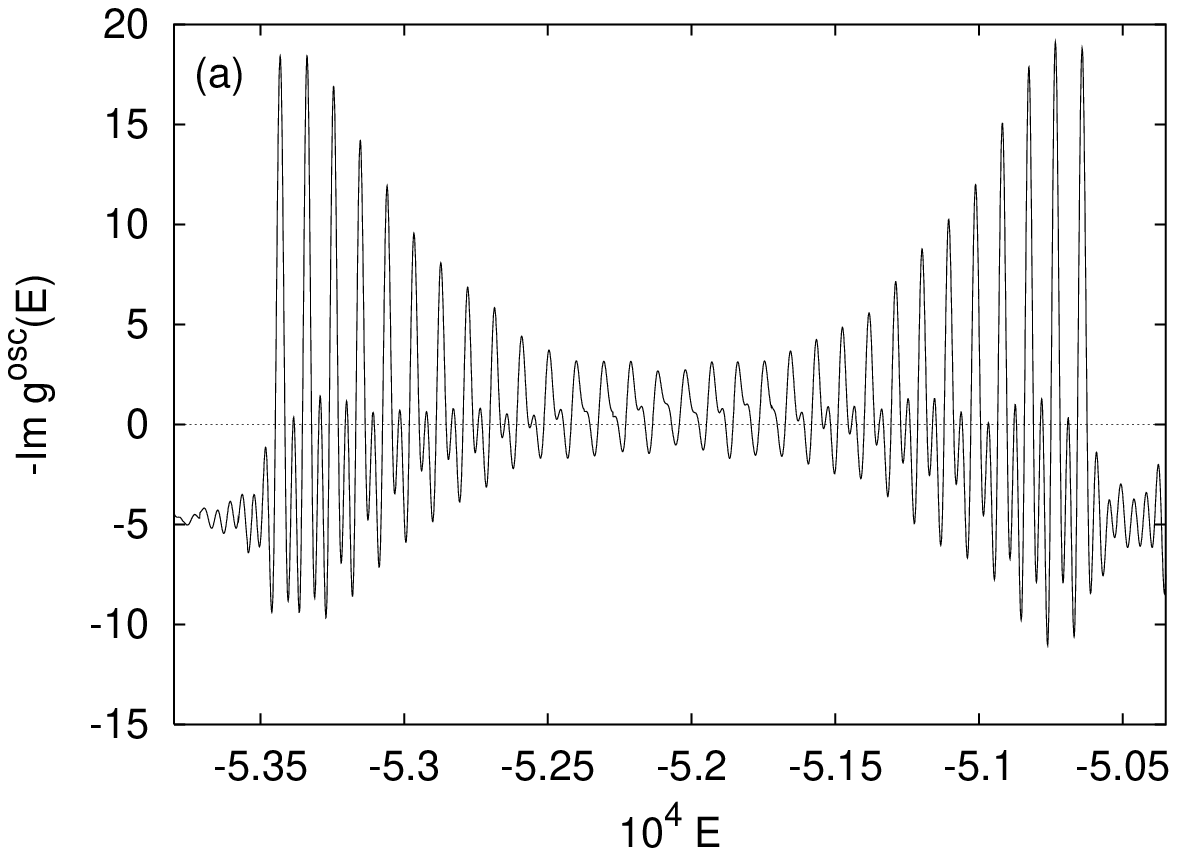}}
  \centerline{\includegraphics[]{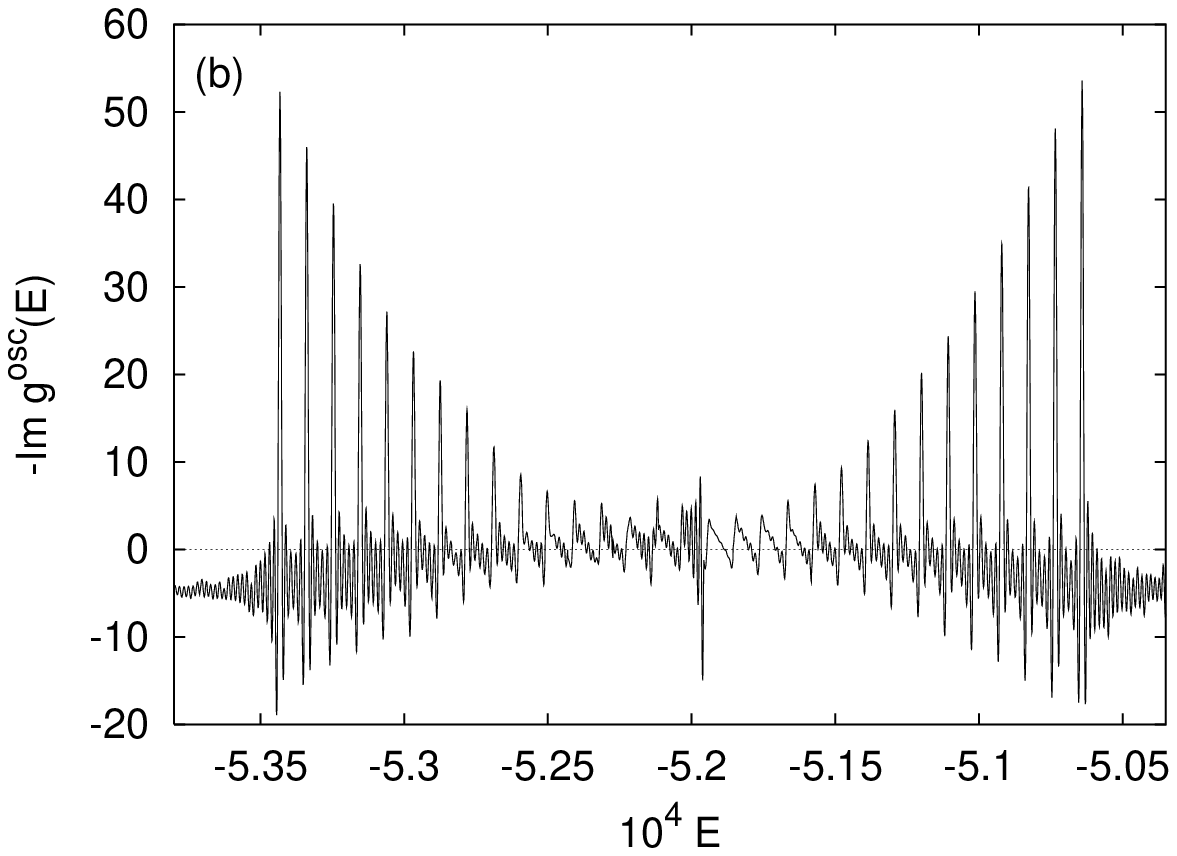}}
  \caption{The manifold $n=31$ in the low-resolution semiclassical spectrum
  with scaled cut-off time (a) $\widetilde T_{\rm max}=15$ and (b) $\tilde
  T_{\rm max}=40$.}
  \label{StarkLoMagFig}
\end{figure}

In the region of isolated $n$-manifolds the quantum and semiclassical
spectra can indeed be compared line by line to reveal excellent agreement
in both the position and the intensity of spectral lines. Note, in
particular, that the low-resolution semiclassical spectrum apparently
ascribes considerable amplitudes to spectral lines in the middle of an
$n$-manifold, whereas the quantum amplitudes nearly vanish. The
high-resolution semiclassical spectrum correctly identifies these
amplitudes as being very small. Furthermore, it can be seen from the
magnified low-resolution spectrum of the manifold $n=31$ in
figure~\ref{StarkLoMagFig}(a) that the spectral lines, in particular the
weak lines close to the centre of the manifold, appear asymmetric in shape
in the semiclassical spectrum: They extend farther to the right than to the
left. Nevertheless, the high-resolution analysis identifies the lines
correctly.

Figure~\ref{StarkLoMagFig}(b) shows the low-resolution spectrum for the
manifold $n=31$ with a signal length of $\widetilde T_{\rm max}=40$. It can
be seen that the spectral lines close to the centre of the manifold become
smaller in comparison to the outer lines, thus approaching the true
semiclassical spectrum, but that the asymmetry of the lines is exacerbated:
They are clearly saw-tooth shaped, rising steeply on the left and falling
off gently to the right.

At $E\approx -4.7\times 10^{-4}$, neighbouring $n$-manifolds overlap for
the first time, thus doubling the density of spectral lines. It can be seen
in figure~\ref{StarkT15HiFig} that at this energy the harmonic inversion of
the given semiclassical signal abruptly breaks down. It recovers at
slightly higher energies, where again only levels of a single $n$-manifold
are present. At $E\approx -4.5\times 10^{-4}$ and $E\approx -4.2\times
10^{-4}$, pairs of levels belonging to different $n$-manifolds are so close
to being degenerate that they cannot be resolved by the harmonic
inversion. In the semiclassical spectrum they appear as single lines with
amplitudes equal to the sum of the two quantum amplitudes.

Similar effects can also be observed at higher energies. However, as the
energy and the density of spectral lines are further increased, the
harmonic inversion gradually ceases to yield meaningful results. In the
high-energy region at $E\approx -3.2\times 10^{-4}$, a few lines can,
somewhat arbitrarily, be identified, whereas most of the lines from the
quantum spectrum are absent. In this region the cut-off time of the
semiclassical signal is evidently, and expectedly, too small.

The harmonic inversion always yields some spurious spectral lines together
with an error parameter $\epsilon$ that can be used to distinguish between
true and spurious lines \cite{Bartsch02}. In figure~\ref{StarkT15HiFig},
only lines with an error parameter $\epsilon<6\times 10^{-8}$ have been
included. This threshold value will also be used for all other
semiclassical spectra presented in this section. The selection of ``good''
semiclassical eigenvalues is therefore solely based on criteria inherent in
the semiclassical quantization procedure, no lines are selected according
to how well they fit the quantum results. As can be seen from
figure~\ref{StarkT15HiFig}, some spurious lines pass the selection. They
are all characterized by having small amplitudes, so that they exert little
influence on the semiclassical signal. Only rarely does it occur that a
true spectral line of considerable amplitude is removed. If the region of
overlapping $n$-manifolds, where the signal is too short to resolve the
spectral lines, is ignored, the only instance of this error present in the
above spectrum can be found at $E\approx -5.1\times 10^{-4}$.

\begin{figure}
  \centerline{\includegraphics[]{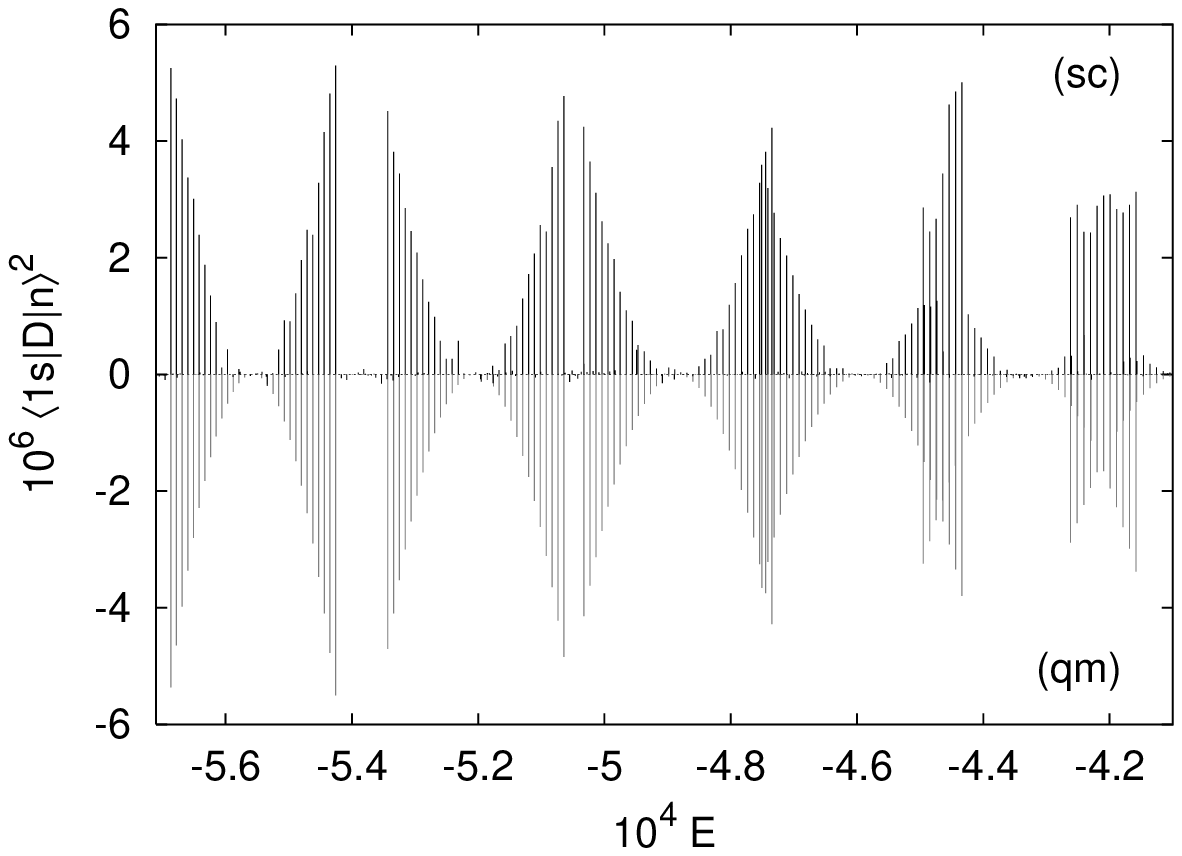}}
  \centerline{\includegraphics[]{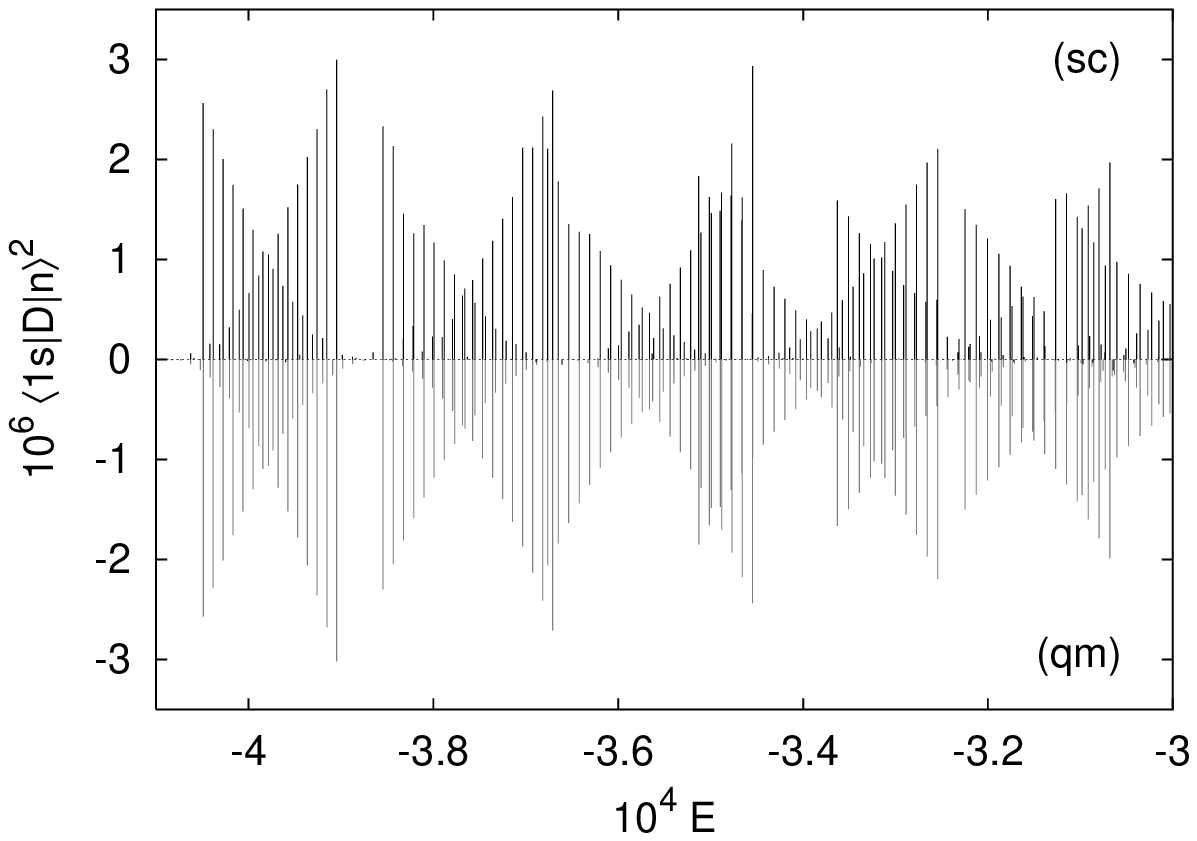}}
  \caption{Same as figure~\ref{StarkT15HiFig} with
   $\widetilde T_{\rm max}=40$.}
  \label{StarkT40HiFig}
\end{figure}

To improve the resolution, the cut-off time needs to be increased. As in
the spectral region around $E\approx -3.2\times 10^{-4}$ three (or even
four) different $n$-manifolds overlap, the true Heisenberg time can be
expected to be close to three times its perturbative
value. Figure~\ref{HeisenbergFig} suggests choosing a scaled cut-off time
of $\widetilde T_{\rm max}=40$. Results obtained with this semiclassical signal
are shown in figure~\ref{StarkT40HiFig}. In the low-energy spectral range,
the amplitudes of the spurious lines have diminished. Thus, even in this
range the longer signal yields a better semiclassical spectrum. In the
region of overlapping $n$-manifolds around $E\approx -3.8\times 10^{-4}$,
all spectral lines are well resolved. At $E\approx -3.2\times 10^{-4}$,
groups of three closely spaced lines can be identified in the
quantum spectrum. In many cases, all three of them are resolved in the
semiclassical spectrum. In a few cases, nearly degenerate lines
appear in the semiclassical spectrum as a single line whose amplitude is
the sum of the two quantum amplitudes.

The most difficult spectral region appears at $E\approx
-4.2\times10^{-4}$. In this region, spectral lines belonging to
neighbouring $n$-manifold are so close to being degenerate that they are
hard to distinguish even in the quantum spectrum. This degeneracy is
somewhat accidental, as it will be lifted as the electric field strength is
varied, but it nevertheless poses a particular challenge to the harmonic
inversion. Even with the signal length of $\widetilde T_{\rm max}=40$, the
degenerate levels cannot be resolved semiclassically. Their resolution
would presumably require a sig\-nifi\-cant\-ly longer semiclassical
signal. As an alternative, the harmonic inversion of cross-correlated
signals has proven powerful in resolving nearly degenerate levels
\cite{Main99c,Weibert00}. It can be combined with the novel quantization
procedure for non-scaling systems in an obvious way and can be expected to
considerably reduce the signal length required to identify the unresolved
spectral lines.

\begin{figure}
  \centerline{\includegraphics[]{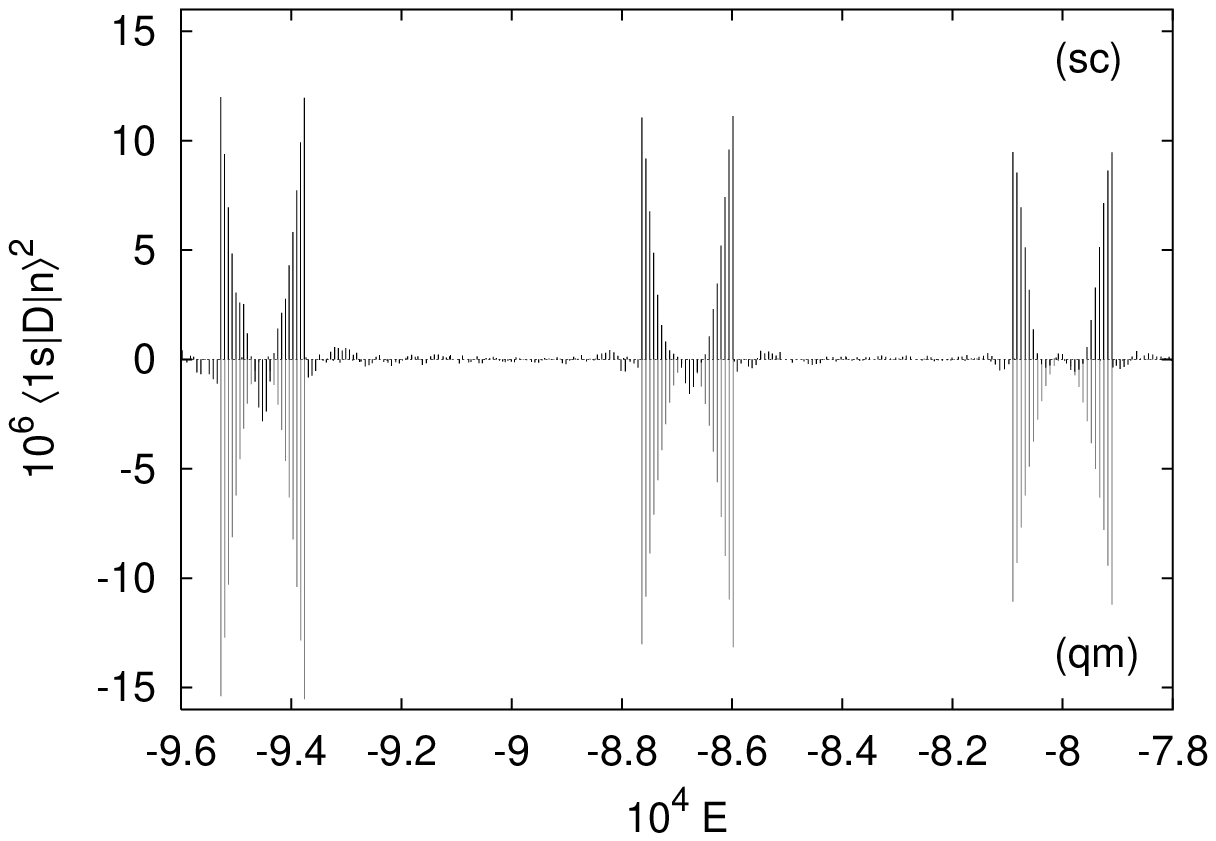}}
  \centerline{\includegraphics[]{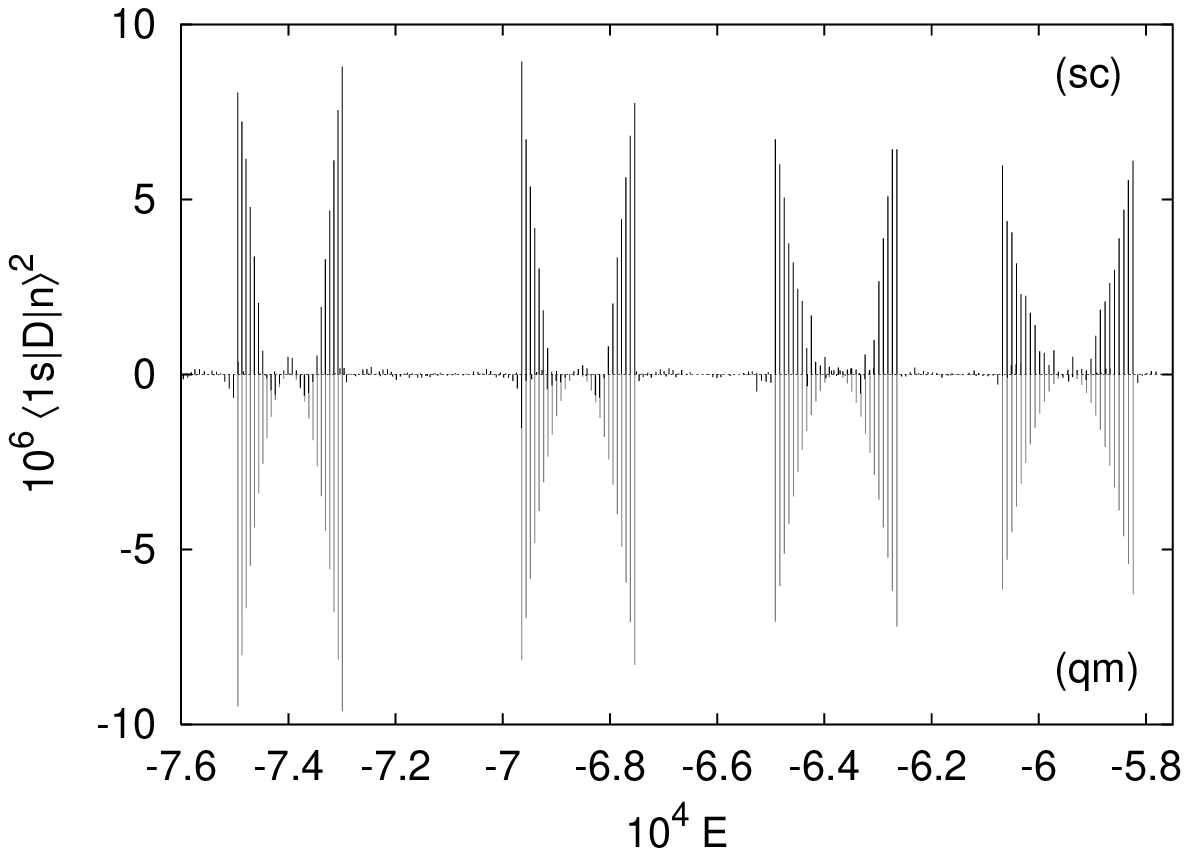}}
  \caption{High-resolution semiclassical (sc) and quantum (qm,
  inverted) photo-absorption spectrum. The scaled cut-off time for
  the semiclassical spectrum is $\widetilde T_{\rm max}=30$.}
  \label{StarkT30HiFig}
\end{figure}

Results obtained with $\widetilde T_{\rm max}=30$ for lower energies are
shown in figure~\ref{StarkT30HiFig}. In this region all $n$-manifolds are
well isolated. As the splitting of levels within a manifold is smaller than
at higher energies, the perturbative Heisenberg time is $\widetilde T_{\rm
H,p}=9.4$ at $E=-10^{-4}$. Therefore, the chosen signal length should
easily suffice to resolve the spectral lines. This is indeed achieved, the
quantum and semiclassical spectra can be compared line by line.  (Most of
the small spurious lines between the manifolds can be removed if a stricter
quality criterion for the semiclassical lines is applied. In
figure~\ref{StarkT30HiFig}, the same threshold $\epsilon<6\times 10^{-8}$
as in previous figures is used.)  However, the agreement between the
spectra is not as perfect as was found at higher energies. In particular,
considerable contributions to the semiclassical spectrum arise at the
centres of the $n$-manifolds at low energies. This decrease in quality
should be expected when the semiclassical approximation is applied to
low-lying states. This is a fundamental limitation of semiclassical
methods. In this case, however, a more subtle difficulty should be
considered: Any non-axial orbits suffers two bifurcations, viz.~its
generations from the downhill orbit and its destruction at the uphill
orbit. At low energies, the distance between the two bifurcations gets
small, so that eventually they can no longer be regarded as isolated. At
this point, the uniform approximation constructed in
section~\ref{sec:StarkUniform} must fail. It must then be replaced with a
uniform approximation describing both bifurcations collectively. Such an
approximation can be constructed, but, contrary to all previous uniform
approximations, it must use a spherical rather than Cartesian configuration
space. This will be discussed in detail in a forthcoming publication
\cite{Bartsch03e}.

An even harder uniformization problem arises close to the Stark saddle
energy. As this energy is approached from below, the downhill orbits
undergo an infinite sequence of bifurcations in a finite energy
interval. Similar bifurcation cascades have been observed in the hydrogen
atom in a magnetic field \cite{Wintgen87} and in H\'enon-Heiles type
systems \cite{Brack01}. Available techniques for the construction of
uniform approximations can only take a finite number of bifurcating orbits
into account, so that they do not allow the uniformization of an infinite
cascade. For this reason, the spectral region around the Stark saddle
energy was excluded from the calculations of the present paper.

\section{Conclusion}

In this paper we have solved the long standing problem of semiclassical
quantization of the hydrogen atom in an electric field based on
closed-orbit theory.  We have presented a detailed study of both the
classical dynamics of the hydrogen atom in an electric field and its
semiclassical description in the framework of closed-orbit theory. The
classical closed orbits were described, explicit formulae for the important
orbital parameters were given, and the pattern of closed-orbit bifurcations
was derived analytically. We constructed a particularly convenient form of
the uniform semiclassical approximation for the closed-orbit bifurcations
in the Stark system. Contrary to earlier versions, our uniform
approximation can be computed from parameters characterizing the isolated
orbits only, so that it is as easy to apply as the simple isolated-orbits
approximation and can immediately be included into a closed-orbit
summation.

We demonstrated that a semiclassical quantization of the hydrogen atom in
an electric field by means of closed-orbit theory is impossible unless
uniform approximations are included.
By means of a generalized method of semiclassical quantization by harmonic
inversion introduced recently \cite{Bartsch02a}, we were able to calculate
high-quality semiclassical photo-absorption spectra from the uniformized
closed-orbit sum. We thus proved that our novel method is indeed well
suited to the extraction of high-resolution semiclassical spectra from
low-resolution spectra of arbitrary systems, whether they are classically
integrable, chaotic, or even mixed regular-chaotic. It thus paves the way
to a semiclassical quantization of the hydrogen atom in a magnetic field or
in crossed electric and magnetic fields. Further work on these more
complicated, classically chaotic systems is in progress. 

The present uniform approximation is applicable only as long as individual
bifurcations can be regarded as isolated. It will fail if they are close.
For this reason, the description of the spectral region close to the Stark
saddle energy remains an open problem. At this energy, the downhill orbits
undergo an infinite cascade of bifurcations in a finite energy
interval. The uniformization of this kind of cascade is far beyond
present-day techniques for the construction of uniform approximations. It
presents a challenge worth-while for future research, both as a problem in
its own right and in view of its potential applications: The bifurcation
cascade occurs at the continuum threshold, where the transition from bound
states to a continuum with embedded resonances takes place
\cite{Kondratovich98}. The semiclassical description of this transition
region is a task of special importance. It will require the uniformization
of the bifurcation cascades.

\section*{References}
%\bibliographystyle{iop}
%\bibliography{../Arbeit}

\end{document}